\RequirePackage{fix-cm}
\documentclass[final]{svjour3}
\smartqed

\usepackage{graphicx}
\usepackage{epsfig}
\usepackage{amsmath}
\usepackage{amssymb}
\usepackage{setspace}
\usepackage{caption}
\usepackage{subcaption}
\usepackage{algpseudocode}
\usepackage[ruled,linesnumbered,vlined]{algorithm2e}
\usepackage{color, colortbl}
\usepackage{array}
\usepackage{hyperref}
\usepackage{multirow}
\usepackage{multicol}
\usepackage{multibib}
\newcolumntype{x}[1]{>{\centering\let\newline\\\arraybackslash\hspace{0pt}}p{#1}}

\definecolor{Gray}{gray}{0.9}

\begin{document}

\title{Landmark-matching Transformation with Large Deformation via $n$-dimensional Quasi-conformal Maps}
\author{Yin Tat Lee \and Ka Chun Lam \and  Lok Ming Lui}
\institute{Yin Tat Lee \at
	Room E18-301A, Department of Mathematics, Massachusetts Institute of Technology \\
	\email{yintat@math.mit.edu}
           \and
           Ka Chun Lam \at
	Department of Mathematics, The Chinese University of Hong Kong \\
	\email{kclam@math.cuhk.edu.hk}
	\and
	Lok Ming Lui \at
	Lady Shaw Building Room 207, Department of Mathematics, The Chinese University of Hong Kong \\	
	Tel: (852) 3943-7975 \\
	\email{lmlui@math.cuhk.edu.hk}
}

\date{Received: date / Accepted: date}

\maketitle

\begin{abstract}
We propose a new method to obtain landmark-matching transformations between n-dimensional Euclidean spaces with large deformations. Given a set of feature correspondences, our algorithm searches for an optimal folding-free mapping that satisfies the prescribed landmark constraints. The standard conformality distortion defined for mappings between 2-dimensional spaces is first generalized to the $n$-dimensional conformality distortion $K(f)$ for a mapping $f$ between $n$-dimensional Euclidean spaces $(n \geq 3)$. We then propose a variational model involving $K(f)$ to tackle the landmark-matching problem in higher dimensional spaces. The generalized conformality term $K(f)$ enforces the bijectivity of the optimized mapping and minimizes its local geometric distortions even with large deformations. Another challenge is the high computational cost of the proposed model. To tackle this, we have also proposed a numerical method to solve the optimization problem more efficiently. Alternating direction method with multiplier (ADMM) is applied to split the optimization problem into two subproblems. Preconditioned conjugate gradient method with multi-grid preconditioner is applied to solve one of the sub-problems, while a fixed-point iteration is proposed to solve another subproblem. Experiments have been carried out on both synthetic examples and lung CT images to compute the diffeomorphic landmark-matching transformation with different landmark constraints. Results show the efficacy of our proposed model to obtain a folding-free landmark-matching transformation between $n$-dimensional spaces with large deformations.

\end{abstract}

\section{Introduction}
\label{intro}
Finding an optimal transformation between corresponding data, such as images or geometric shapes, is an important task in various fields, such as computer visions \cite{SurveyZitova}, computer graphics \cite{la2000fast,Haker,TextureMapSurvey}, video processing \cite{LuiBeltramirepresentation,reddy1996fft,Weiface,zhu2000new}  and medical imaging \cite{avants2008symmetric,Hill,Joshi,klein2010elastix,LamMICCAI,DMIRsurvey,vannier1996three}. Such a process is called {\it registration}. For example, in neuroimaging, it is often required to align medical images from different modalities, such as magnetic resonance (MR), X-ray computed tomography (CT) images and so on. In computer graphics, registration is necessary for texture mapping \cite{Haker,TextureMapSurvey}. Due to its important applications in different areas, an enormous amount of research has been carried out to develop effective models for registration.

Registration methods can mainly be divided into three categories, namely, 1. intensity-based registration, 2. landmark-based registration and 3. hybrid registration using both intensity and landmark information. Intensity-based registration computes a transformation between corresponding data by matching intensity functions, such as image intensity for image registration or surface curvature for surface geometric registration. Different intensity-based registration algorithms have been recently proposed \cite{SurveyZitova}, such as Demons \cite{Thirion,Vercauteren}, spherical Demons \cite{Yeo}, elastic registration \cite{he2003large}, Large Deformation diffeomorphic Metric Mapping (LDDMM) frameworks \cite{christensen1996deformable,dupuis1998variational} and so on. On the other hand, landmark-based registration computes a smooth 1-1 dense pointwise correspondence between corresponding data that matches important features \cite{Bookstein,Glaunes3,Glaunes2,Glaunes,Joshi,lui2008optimized,Lui10,Luilandmark,Tosun,Glaunes4,Wang05}. Such a feature-based registration approach usually comprises of two steps, namely, 1. the extraction of corresponding feature landmarks and 2. the computation of a transformation between the data that matches corresponding features. The main advantage of the landmark-based method is that intuitive user-interaction can be incorporated to guide the registration process. Recently, hybrid registration that combines landmark-based and intensity-based methods have also gained increased attention. Hybrid approaches use both the landmark and intensity information to guide the registration. This type of approaches can usually obtain more accurate registration results, since the advantages of landmark-based and intensity-based registration can be combined. Different hybrid registration models have also been proposed recently \cite{Hybrid2,DROP,HybridHuang,Johnson02consistentlandmark,Hybrid1}.

In this work, we will focus on the landmark-based registration. Landmark-based registration has found important applications. One typical example is the brain cortical surface registration for which sulcal landmarks are usually extracted to guide the registration \cite{Luilandmark,Tosun,Wang05}. Landmark-based registration has also been applied to register gene expression data to a neuroanatomical mouse atlas \cite{Lin}. Feature-matching image registration can also be used as an initial guess for intensity-based registration between images with large deformations \cite{Johnson02consistentlandmark,LuiQFibra}. Over the past few decades, numerous landmark-based registration models have been proposed \cite{Bookstein,guo2006diffeomorphic,Joshi,kybic2003fast,rohr2001landmark}. One of the first and most important landmark-based registration algorithm is the Thin-Plate Spline (TPS) method proposed by Bookstein \cite{Bookstein}. TPS minimizes the bending energy together with the landmark mismatching term. A unique and closed-form solution is guaranteed in this model. TPS is efficient and works well under small deformations. However, under larger deformations, TPS generally cannot preserve the bijectivity of the mapping \cite{TPSfold}. 

In some situations like medical image registration or constrained texture mapping of surfaces, a bijective and topology-preserving mapping is desirable for the registration problem \cite{DMIRsurvey}. For example, Christensen et. al \cite{christensen1996deformable} proposed a regridding algorithm to restrict the transformation of the image deformation to have a globally positive definite Jacobian. Statistically, Leow et. al \cite{leow2007statistical} studied the statistical properties of Jacobian maps (the determinant of the Jacobian matrix of a deformation field) and proposed a framework for constructing unbiased deformation fields. Modat et.al \cite{modat2010lung} also proposed a variational model with the joint bending energy and the squared Jacobian determinant penalty terms to obtain a transformation for lung registration.

Recently, quasi-conformal (QC) theory has been introduced to handle large deformation landmark-matching registration problem \cite{Lipman3,LuiQFibra,Lipman1,Weiface}. The Beltrami coefficient, which measures the conformality distortion, can be effectively used to enforce the bijectivity of the mapping. By optimizing an energy functional involving the $L^p$-norm of the Beltrami coefficient, large deformation diffeomorphic registration can be accurately computed. Several works have also been proposed to deal with surface-based landmark-matching problem with different genus \cite{LamMICCAI,LuiHG}. QC theory has provided an effective framework to handle registration problem with large deformations for 2-dimensional spaces. However, for general $n$-dimensional spaces, the conformality distortion is not defined. Motivated by this, it is our goal in this paper to extend the concept of 2D quasi-conformality to general $n$-dimensional spaces. In particular, a notion of conformality distortion of a diffeomorphism in the $n$-dimensional Euclidean space will be formulated. With the definition of conformality distortion, we can extend the 2D quasi-conformal registration algorithm to general $n$-dimensional Euclidean spaces. 

In short, the main contributions in this paper are three-folded:
\begin{enumerate}
\item We give a definition of $n$-dimensional conformality distortion $K(f)$ ($n\geq 3$) for mapping $f$ between $n$-dimensional Euclidean spaces. The standard conformality distortion is defined for 2-dimensional space. Our definition aims to generalize this concept to $n$-dimensional spaces.

\item With the definition of $K(f)$, we extend our previous model \cite{LuiQFibra} for computing 2-dimensional landmark-matching bijective mapping with large deformations to higher dimensional spaces. This allows us to compute bijective landmark-matching mapping of higher dimensional spaces with large deformations.

\item One challenge of the proposed model is the high computational cost for higher-dimensional spaces. In this paper, we propose a numerical method to solve the optimization problem more efficiently. This is based on applying the alternating direction method with multiplier (ADMM) to split the problem into two subproblems. Preconditioned conjugate gradient method with multi-grid V-cycle preconditioner is applied to solve one of the subproblems. A fixed-point iteration is proposed to solve another subproblem, whose convergence to the minimizer is theoretically shown.
\end{enumerate}

This paper is organized as follows. In section 2, basic mathematical background will be explained. In section 3, we describe our proposed model to obtain the landmark-matching transformation with large deformation between $n$-dimensional Euclidean spaces in details. The numerical algorithm will be discussed in section 4. Experimental results will be demonstrated in section 5. Conclusion and future work will be discussed in section 6.

\section{Mathematical background}
\label{sec:2}
In this section, we describe some basic mathematical concepts related to our algorithms. For details, we refer the readers to \cite{Gardiner,Lehto}.

A surface $S$ with a conformal structure is called a \emph{Riemann surface}. Given two Riemann surfaces $M$ and $N$, a map $f:M\to N$ is \emph{conformal} if it preserves the surface metric up to a scalar multiplicative factor called the conformal factor. An immediate consequence is that every conformal map preserves angles. With the angle-preserving property, a conformal map effectively preserves the local geometry of the surface structure. 

A generalization of conformal maps is the \emph{quasi-conformal} maps, which are orientation preserving homeomorphisms between Riemann surfaces with bounded conformality distortion, in the sense that their first order approximations takes small circles to small ellipses of bounded eccentricity \cite{Gardiner}. Surface registrations and parameterizations, which are orientation-preserving homeomorphisms, can be considered as quasi-conformal maps. Mathematically, $f \colon \mathbb{C} \to \mathbb{C}$ is quasi-conformal provided that it satisfies the Beltrami equation:
\begin{equation}\label{beltramieqt}
\frac{\partial f}{\partial \overline{z}} = \mu(z) \frac{\partial f}{\partial z}.
\end{equation}
\noindent for some complex valued Lebesgue measurable $\mu$ satisfying $||\mu||_{\infty}< 1$. $\mu$ is called the \emph{Beltrami coefficient}, which is a measure of non-conformality. In particular, the map $f$ is conformal around a small neighborhood of $p$ when $\mu(p) = 0$. Infinitesimally, around a point $p$, $f$ may be expressed with respect to its local parameter as follows:
\begin{equation}
\begin{split}
f(z) & \approx f(p) + f_{z}(p)z + f_{\overline{z}}(p)\overline{z} \\
& = f(p) + f_{z}(p)(z + \mu(p)\overline{z}).
\end{split}
\end{equation}

Obviously, $f$ is not conformal if and only if $\mu(p)\neq 0$. Inside the local parameter domain, $f$ may be considered as a map composed of a translation to $f(p)$ together with a stretch map $S(z)=z + \mu(p)\overline{z}$, which is postcomposed by a multiplication of $f_z(p),$ which is conformal. All the conformality distortion of $S(z)$ is caused by $\mu(p)$. $S(z)$ is the map that causes $f$ to map a small circle to a small ellipse. From $\mu(p)$, we can determine the angles of the directions of maximal magnification and shrinking and the amount of them as well. Specifically, the angle of maximal magnification is $\arg(\mu(p))/2$ with magnifying factor $1+|\mu(p)|$; The angle of maximal shrinking is the orthogonal angle $(\arg(\mu(p)) -\pi)/2$ with shrinking factor $1-|\mu(p)|$. The distortion or dilation is given by:
\begin{equation}
K = \left({1+|\mu(p)|}\right)/\left({1-|\mu(p)|}\right).
\end{equation}

\begin{figure}[t]
        \centering
        \begin{subfigure}[b]{0.4\textwidth}
                \includegraphics[width=\textwidth]{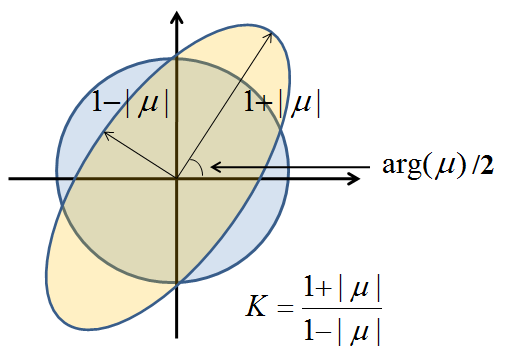}
                \caption{}
                \label{fig:conformalitydistortionA}
        \end{subfigure}
        \begin{subfigure}[b]{0.45\textwidth}
                \includegraphics[width=\textwidth]{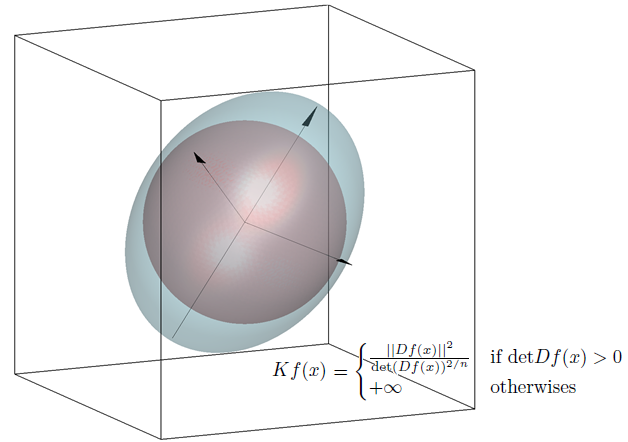}
                \caption{}
                \label{fig:conformalitydistortionB}
        \end{subfigure}
\caption{Illustration of the conformality distortion. (a) shows how a small circle is deformed to an ellipse under a 2D quasi-conformal map. The conformality distortion is measured by the Beltrami coefficient. (b) shows how a small ball is deformed to a small ellipsoid under a 3D diffeomorphism. The conformality distortion can be measured by $K(f)$ defined in this paper.}
\end{figure}

Thus, the Beltrami coefficient $\mu$ gives us all the information about the properties of the map (See Figure \ref{fig:conformalitydistortionA}).

Given a Beltrami coefficient $\mu:\mathbb{C}\to \mathbb{C}$ with $\|\mu\|_\infty < 1$. There is always a quasi-conformal mapping from $\mathbb{C}$ onto itself which satisfies the Beltrami equation in the distribution sense \cite{Gardiner}.

However, the above quasi-conformal theories only apply to two dimensional spaces or surfaces. In this work, our goal is to extend the idea of 2-dimensional quasi-conformal theories to general $n$-dimensional spaces. We will introduce a notion of conformality distortion of a diffeomorphism of the $n$-dimensional Euclidean space. The conformality distortion measures the distortion of an infinitesimal ball to an infinitesimal ellipsoid under the diffeomorphism (See Figure \ref{fig:conformalitydistortionB}).

\section{Proposed model}
\label{sec:3}
In this section, we will explain in details our proposed model to obtain the landmark-matching transformation between $n$-dimensional Euclidean spaces. The basic idea is to formulate the notion of conformality distortion of a diffeomorphism of the $n$-dimensional Euclidean space. The conformality distortion measures the distortion of an infinitesimal ball to an infinitesimal ellipsoid under the diffeomorphism. The landmark-matching problem can then be modelled as minimizing an energy functional involving a conformality term and a smoothness term under the prescribed landmark constraints. We first introduce the conformality distortion of a diffeomorphism of the n-dimensional Euclidean space. In subsection 3.2, we describe the continuous model of the proposed energy functional. Finally, we explain the discretization of the model in subsection 3.3.

\subsection{Conformality distortion}
\label{sec:3.1}
Let $\Omega_1, \Omega_2 \subset \mathbb{R}^{n}$ be the domain and the image of the diffeomorphism $f=(f_1,f_2,...,f_n) : \Omega_1 \rightarrow \Omega_2$ respectively. For any $\mathbf{p} = (p_1,p_2,...,p_n) \in \Omega_1$, let $\mathbf{q} = (q_1,q_2,...,q_n) = f(\mathbf{p})$. Then, for any $\mathbf{x} = (x_1,...,x_n)$ in a neighbourhood of $\mathbf{p}$, we have
\begin{equation}
\mathbf{y}= f(\mathbf{x}) \approx f(\mathbf{p}) + Df (\mathbf{x}-\mathbf{p}),
\end{equation}
\noindent where $Df = (\frac{\partial f_i}{\partial x_j})_{1 \leq i,j\leq n} \in M_{n\times n}(\mathbb{R})$.

Under a general diffeomorphism $f$, $f$ distorts an infinitesimal ball $B_{\epsilon}(\mathbf{p}) := \{\mathbf{x}\in \Omega_1: ||\mathbf{x}-\mathbf{p}||\leq \epsilon\}$ to an infinitesimal ellipsoid $E_f$ (see Figure \ref{fig:conformalitydistortionB}). More precisely,
\begin{equation}
\begin{split}
E_f &= \{ \mathbf{q} + Df (\mathbf{x}-\mathbf{p}): \mathbf{x}\in B_{\epsilon}(\mathbf{p}) \}\\
& = \{ \mathbf{q} + \mathbf{w}: \mathbf{w}^T C \mathbf{w} \leq \epsilon \},
\end{split}
\end{equation}
\noindent where $C = ((Df)^{-1})^T(Df)^{-1}$ is a symmetric positive definite matrix. Obviously, since $C$ is symmetric positive definite, $E_f$ is an ellipsoid centered at $\mathbf{q}$. Moreover, $E_f$ is a infinitesimal ball if all eigenvalues of $C$ are equal. This can be observed easily as follows. Suppose $C = Q^TD Q$, where $Q$ is an orthogonal matrix and $D$ is a diagonal matrix consisting of the eigenvalues of $A$. If $D = \mu \mathbf{I}$ ($\mu>0$), $C = \mu\mathbf{I}$. It follows that for any $\mathbf{y} \in E_f$, $(\mathbf{y}-\mathbf{q})^T C(\mathbf{y}-\mathbf{q}) = \mu (\mathbf{y}-\mathbf{q})^T (\mathbf{y}-\mathbf{q})\leq \epsilon$. This gives $(\mathbf{y}-\mathbf{q})^T (\mathbf{y}-\mathbf{q})\leq \epsilon/\mu$ Hence, $E_f$ is an infinitesimal ball with radius $\epsilon/\mu$.

To define the conformality distortion, we define a measurement that quantifies the geometric distortion of the ellipsoid $E_f$ from an infinitesimal ball. From the above observation, it is the same as measuring how far the matrix $C$ is from a symmetric positive definite matrix with equal eigenvalues. It is related to the Jacobian of the mapping $f$.

Consider $A = (Df)^T (Df)$. The eigenvalues of $C$ are equal if and only if the eigenvalues of $A=(Df)^T (Df)$ are equal. Suppose $\lambda_1,\lambda_2,...,\lambda_n$ are the eigenvalues of $A$. Using the AM--GM inequality, we have
\begin{equation}
\begin{split}
\left(\lambda_{1}\cdots\lambda_{n}\right)^{1/n} & \leq\frac{\lambda_{1}+\cdots+\lambda_{n}}{n} \ \ \text{ or},\\
\frac{1}{n} \left( \frac{\lambda_{1}+\cdots+\lambda_{n}}{\left(\lambda_{1}\cdots\lambda_{n}\right)^{1/n}} \right) & \geq 1
\end{split}
\end{equation}
\noindent where the equality sign holds if and only if $\lambda_{1}=\cdots=\lambda_{n}$. Hence, we can define the $n$-D conformality distortion as $K(f) = \frac{1}{n} \left( \frac{\lambda_{1}+\cdots+\lambda_{n}}{\left(\lambda_{1}\cdots\lambda_{n}\right)^{1/n}} \right)$. Note that the AM-GM inequality states that the perimeter of the $n$-dimensional cubes is the smallest amongst all $n$-dimensional rectangular boxes with the same volume. In the 2-dimensional case, the perimeter of a square is always the smallest amongst all rectangles with a given area. Hence, $K(f)$ can be interpreted as the ratio of the perimeter of a $n$-dimensional rectangular boxes with edges lengths equal to the eigenvalues of $A$ to the perimeter of the $n$-cube with the same volume. Let $\{\mathbf{v}_1,...,\mathbf{v}_n\}$ be the orthonormal basis of eigenvectors of $A$. Then, the minimum is attained if the $n$-dimensional boxes spanned by $\{\mathbf{v}_1,...,\mathbf{v}_n\}$ is a $n$-cube. This happens when all eigenvalues are equal.

Now, the arithmetic mean and geometric mean of the eigenvalues of $A$ can be expressed as the Frobenius norm and determinant of $Df$ respectively. Observe that:
\begin{equation}
\begin{split}
||Df||_F^2 &= \mathrm{Tr}(Df^T Df) = \mathrm{Tr}(A) = \lambda_{1}+\cdots+\lambda_{n};\\
\mathrm{det}(A) &=\lambda_{1}\cdots\lambda_{n}= \mathrm{det}(Df^T Df) = \mathrm{det}(Df)^2.
\end{split}
\end{equation}

Therefore, we can now introduce the following definition:

\begin{definition}[Conformality distortion]
\label{nd_mu}
The conformality distortion $Kf(x)$ of a mapping $f$ at point $x$ is defined by
\begin{eqnarray}
Kf(x) & := & \begin{cases}
\frac{1}{n} \left( \frac{||Df(x)||_{F}^{2}}{\det(Df(x))^{2/n}}\right)  & \text{if }\text{det}\left( Df(x) \right)>0,\\
\text{+\ensuremath{\infty}} & \text{otherwise}
\end{cases}
\end{eqnarray}
where $ ||Df(x)||_{F}^{2} = \text{Tr}(Df(x)^{T}Df(x))$ denotes the Frobenius norm of $Df(x)$.
\end{definition}
\noindent Note that $Kf(x)\geq 1$ and $Kf(x)=1$ if and only if $E_{\mathbf{x}}= \{ \mathbf{x} + Df (\mathbf{y}-\mathbf{x}): \mathbf{y}\in B_{\epsilon}(\mathbf{x}) \}$ is a $n$-dimensional ball. Motivated by this observation,  we say $f$ is conformal at point $x$ if the conformality distortion $Kf(x)$ attains its minimum value $1$. By setting $Kf(x) = + \infty$ when $\text{det}(Df(x)) \leq 0$, we can ensure the bijectivity of the mapping by minimizing the norm of $Kf(x)$.

For $n = 2$, denote $f(x_1,x_2) = f_1(x_1,x_2) + \sqrt{-1} f_2(x_1,x_2)$ and assume $\text{det}(Df(x)) > 0$ for all $x$. Then, $\frac{\partial f}{\partial \overline{z}} = (\frac{\partial f_1}{\partial x_1} + \frac{\partial f_2}{\partial x_2}) + \sqrt{-1}(\frac{\partial f_2}{\partial x_1} - \frac{\partial f_1}{\partial x_2})$ and $\frac{\partial f}{\partial z} = (\frac{\partial f_1}{\partial x_1} - \frac{\partial f_2}{\partial x_2}) + \sqrt{-1}(\frac{\partial f_2}{\partial x_1} + \frac{\partial f_1}{\partial x_2})$. We have
\begin{equation}
\begin{split}
Kf(x) = \frac{1}{2}\frac{\Vert Df(x)\Vert_{F}^2}{\text{det}(Df(x))} &= \frac{\left(\frac{\partial f_1}{\partial x_1}\right)^2 +\left(\frac{\partial f_1}{\partial x_2}\right)^2 + \left(\frac{\partial f_2}{\partial x_1}\right)^2 + \left(\frac{\partial f_2}{\partial x_2}\right)^2}{2\text{det}(Df(x))} \\
&= \frac{ \vert f_z \vert^2(1 + \vert \mu \vert^2)}{\vert f_z \vert^2 (1 - \vert \mu \vert^2)} \\
&= \frac{1+ \vert \mu \vert^2}{1 - \vert \mu \vert^2}
\end{split}
\end{equation}
\noindent where $\mu(x)$ is the Beltrami coefficient defined in equation (\ref{beltramieqt}).

\subsection{The continuous model}
\label{sec:3.2}

With the notion of n-D conformality distortion $K(f)$, we can now develop a variational model to compute a landmark-matching transformation between n-dimensional spaces. Given two domains $\Omega_1$ and $\Omega_2$ in $\mathbb{R}^n$.  Suppose $\{p_i \in \Omega_1\}_{i=1}^m$ $\{q_i \in \Omega_2\}_{i=1}^m$ are corresponding feature landmarks in $\Omega_1$ and $\Omega_2$ respectively. These corresponding sets of feature landmarks gives the landmark constraints on the mapping. Our goal is to search for a bijective transformation $f:\Omega_1\to \Omega_2$ that satisfies $f(p_i) = q_i$ for $i=1,2,..., m$. In other words, the selected landmark points $p_i$ in $\Omega_1$ are required to mapped to the corresponding $q_i$ in $\Omega_2$. This is called the landmark-based registration problem. Most of the existing landmark-based registration models are variational approaches. They can mainly be written as minimizing:
\begin{equation}
E(f) = {\it Reg}(f) \text{ subject to: } f(p_i) = q_i \text{ for } i=1,...,n \text{ (hard landmark constraints),}
\end{equation}
\noindent where $Reg(f)$ is the regularization of the mapping $f$. Sometimes, the hard landmark constraints can be relaxed by minimizing
\begin{equation}
E(f) = {\it Reg}(f) + \lambda \sum_{i=1}^n ||f(p_i) - q_i||_2^2 \text{ (soft landmark constraints)}
\end{equation}


For example, the popular thin-plate spline (TPS) landmark-based registration model uses the integral of the square of the second derivative as the regularizer \cite{Bookstein}. In this work, we propose to use the $L^1$-norm of the conformality distortion $K(f)$ together with a smoothness term $\Vert \Delta f \Vert^2_2$ as the regularizer. Note that the conformality distortion has also been used to obtain registration for 2-dimensional spaces (such as 2D images or 2D surfaces) \cite{LuiQFibra}. In this paper, with the notion of $n$-D conformality distortion $Kf(x)$, we extend this idea to general $n$-dimensional spaces. This allows us to compute bijective landmark-matching mappings of higher dimensional spaces with large deformations. Another challenge is the high computational cost of the proposed model in the high dimensional space. To handle this, we also propose a numerical method to solve the optimization problem more efficiently (please refer to Section \ref{sec:4}).

With conformality distortion $Kf(x)$, the bijectivity of the registration can be easily guaranteed by enforcing the constraint $||Kf(x)||_{\infty}<K$ for some $K<\infty$. This can be achieved by minimizing an energy functional involving $||Kf(x)||_{\infty}$. In addition, minimizing $||Kf(x)||_{\infty}$ also helps to reduce the maximal conformality distortion, and hence reduce the local geometric distortion of the mapping. However, it is computationally expensive to minimize an energy functional involving the supremum norm. Consequently, we propose to minimize $||Kf(x)||_{1}$. Since $Kf(x)$ is set to be $+ \infty$ when $\text{det}(Df(x)) \leq 0$, our variational model can still prevent folding by minimizing $||Kf(x)||_{1}$.

Besides, $\Vert Df \Vert_{F}$ is included in the term $Kf$, which is the commonly used squared Frobenius regularization term. The smoothness of the mapping can be achieved by merely minimizing the conformality term. In order to further enhance the smoothness of the mapping, an extra smoothness term can be included in the energy functional. We now propose the minimization model for the landmark-based registration problem as follows:
\begin{equation}
\inf_{f\in F}\Vert Kf(x) \Vert_1+\frac{\sigma}{2}||\Delta f(x)||_{2}^{2}dx\label{eq:our_energy}
\end{equation}
\noindent where $\sigma\geq 0$ is a fixed parameter and $F = \left\{ f:\Omega \subset \mathbb{R}^n \rightarrow \mathbb{R}^n \vert f(p_i) = q_i, i=1,2,\ldots m \right\}$ is the set of functions $f:\Omega\rightarrow\mathbb{R}^{n}$ which satisfies
the landmark constraint $f(p_{i})=q_{i}$, where $p_{i}$ and $q_{i}$ are the given landmark points ($i=1,2,\ldots ,m$). The first energy term helps to obtain a quasi-conformal map with minimal conformality distortion, while satisfying the landmark constraints. The second energy term aims to further enhance the smoothness of the mapping, since it involves higher order derivatives. Again, since $\Vert Df \Vert_{F}$ is included in $Kf$, some smoothness can already be achieved by minimizing the first energy term. In practice, we set $\sigma = 0$, which is enough to give smooth landmark-aligned mappings. This improves the efficiency of the algorithm. In extreme situations (such as a very large deformation), setting a non-zero $\sigma$ can help to achieve much smoother registration results.
\subsection{The discrete model}
\label{sec:4.3}
For general Euclidean space, our model (\ref{eq:our_energy}) can be discretized by using discrete differential forms. For the ease of explanation, we will explain the discretization of (\ref{eq:our_energy}) on a cubic domain in the 3-D space here. First, we pick a tetrahedral mesh for the cubic domain such that each tetrahedron in the mesh contains 3 edges, each one of them is parallel to the one of the three coordinate axis respectively. In our implementation, we partition the cubic domain into small equal-size cubes and create similar tetrahedral meshes for each cubes. For the unit cube with vertices $\{ {\bf x}_1, {\bf x}_2, {\bf x}_3, {\bf x}_4, {\bf x}_5, {\bf x}_6, {\bf x}_7, {\bf x}_8\}$
we use the tetrahedral mesh with 6 tetrahedra. The vertices for these 6 tetrahedra are:
\begin{figure}[!h]
\center
\begin{minipage}{.22\textwidth}
\vspace{10pt}
\includegraphics[width=\linewidth]{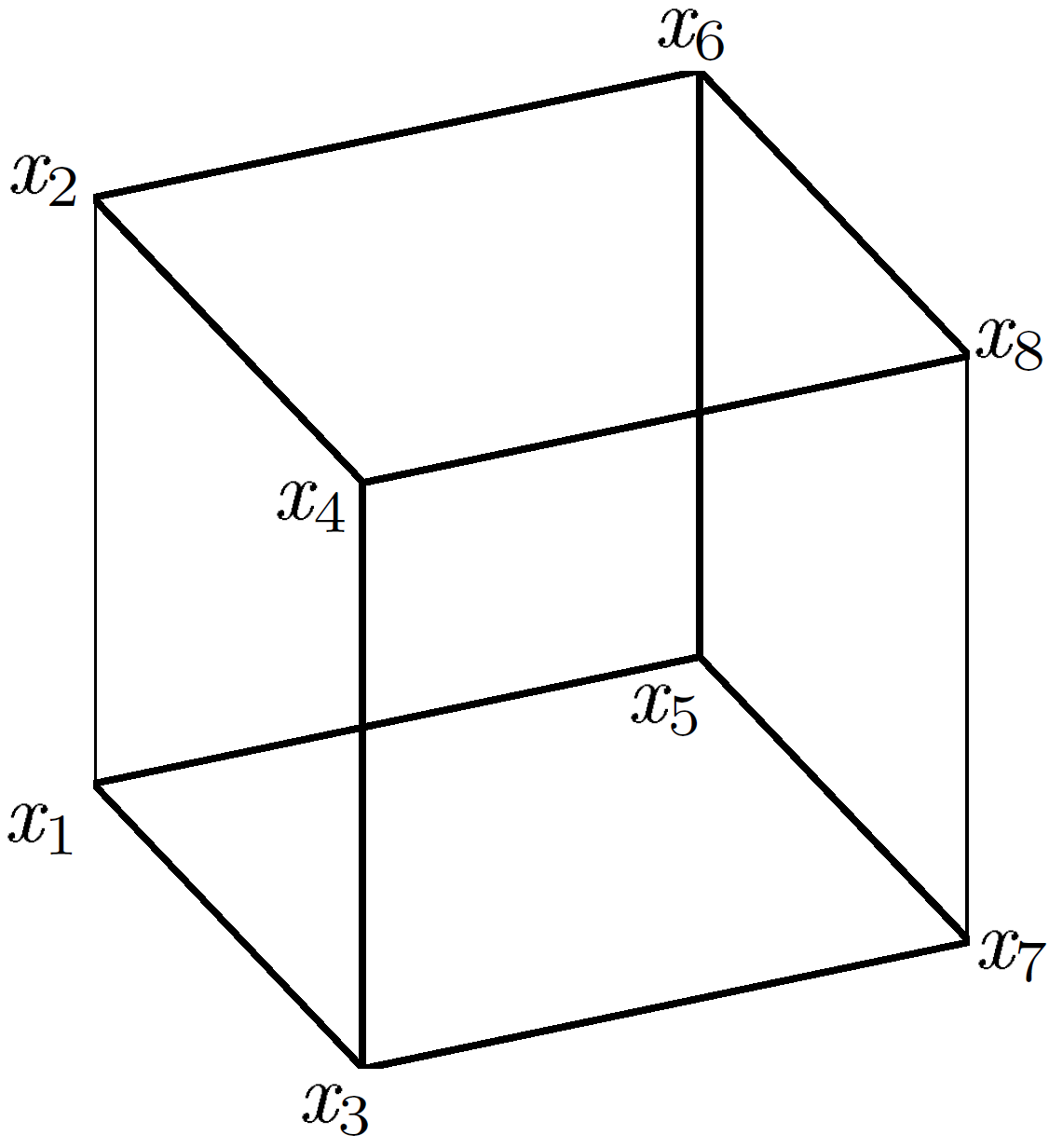}
\end{minipage}
\begin{minipage}{.3\textwidth}
\vspace{0pt}
\begin{eqnarray*}
 1 :& \{{\bf x}_3, {\bf x}_7, {\bf x}_4, {\bf x}_5\} & ,\\
 2 :& \{{\bf x}_3, {\bf x}_1, {\bf x}_4, {\bf x}_5\} & ,\\
 3 :& \{{\bf x}_4, {\bf x}_1, {\bf x}_2, {\bf x}_5\} & ,\\
 4 :& \{{\bf x}_7, {\bf x}_4, {\bf x}_5, {\bf x}_8\} & ,\\
 5 :& \{{\bf x}_4, {\bf x}_5, {\bf x}_8, {\bf x}_6\} & ,\\
 6 :& \{{\bf x}_4, {\bf x}_2, {\bf x}_5, {\bf x}_6\} & .
\end{eqnarray*}
\end{minipage}
\begin{minipage}{.24\textwidth}
\vspace{10pt}
\includegraphics[width=\linewidth]{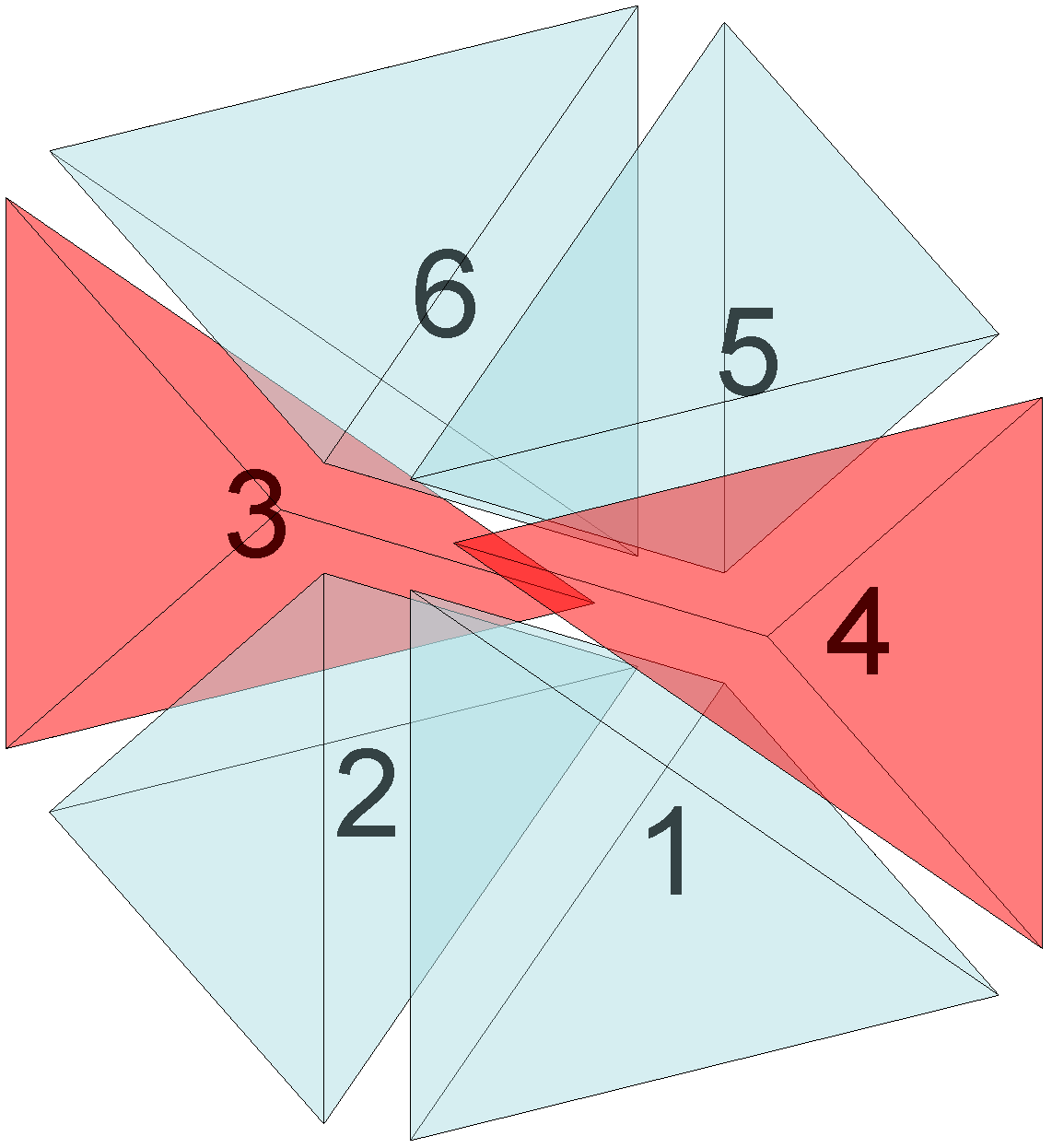}
\end{minipage}
\end{figure}

Consider the affine map $A$ associated with each tetrahedron. Denote ${\bf u}_k = (x_k,y_k,z_k) \in \mathbb{R}^3, \quad k=0,1,2,3$ be the coordinates of the four vertices of the tetrahedron in Euclidean space. We also denote the image of the affine map to be $A({\bf u}_k) = {\bf v}_k = (\tilde{x}_k, \tilde{y}_k, \tilde{z}_k) \in \mathbb{R}^3$. In matrix notation, we have
\begin{equation}
\left( \begin{array}{cccc} \tilde{x}_0 & \tilde{x}_1 & \tilde{x}_2 & \tilde{x}_3 \\ \tilde{y}_0 & \tilde{y}_1 & \tilde{y}_2 & \tilde{y}_3 \\ \tilde{z}_0 & \tilde{z}_1 & \tilde{z}_2 & \tilde{z}_3 \\ 1 & 1 & 1 & 1 \end{array} \right) = A \left( \begin{array}{cccc} x_0 & x_1 & x_2 & x_3 \\ y_0 & y_1 & y_2 & y_3 \\ z_0 & z_1 & z_2 & z_3 \\ 1 & 1 & 1 & 1 \end{array} \right), \quad \text{ where } A = \left( \begin{array}{cccc} a_{00} & a_{01} & a_{02} & a_{03} \\ a_{10} & a_{11} & a_{12} & a_{13} \\ a_{20} & a_{21} & a_{22} & a_{23} \\ a_{30} & a_{31} & a_{32} & a_{33} \end{array} \right)
\end{equation}
We then have
\begin{equation}
A = \left( \begin{array}{cccc} \tilde{x}_0 & \tilde{x}_1 & \tilde{x}_2 & \tilde{x}_3 \\ \tilde{y}_0 & \tilde{y}_1 & \tilde{y}_2 & \tilde{y}_3 \\ \tilde{z}_0 & \tilde{z}_1 & \tilde{z}_2 & \tilde{z}_3 \\ 1 & 1 & 1 & 1 \end{array} \right) \left( \begin{array}{cccc} x_0 & x_1 & x_2 & x_3 \\ y_0 & y_1 & y_2 & y_3 \\ z_0 & z_1 & z_2 & z_3 \\ 1 & 1 & 1 & 1 \end{array} \right)^{-1}
\end{equation}
Thus the Jacobian matrix of the affine map $A$ is 
\begin{equation}
D(A) = \left( \begin{array}{ccc} a_{00} & a_{01} & a_{02} \\ a_{10} & a_{11} & a_{12} \\ a_{20} & a_{21} & a_{22} \end{array} \right)
\end{equation}
Denote $Df(T)$ to be the $3\times3$ Jacobian matrix of $f$ for tetrahedron $T$. The discrete version of (\ref{eq:our_energy}) is given by
\begin{equation}\label{discreteenergy}
\inf_{f\in F}\sum_{\text{tetrahedron }T}Kf(T)+\frac{\sigma}{2}\sum_{\text{node }x}||\Delta f(x)||_2^{2}
\end{equation}
where $f \in F = \left\{ f : \Omega \subset \mathbb{R}^3 \rightarrow \mathbb{R}^3 \vert f(p_i) = q_i,\ i = 1,2,\ldots,m\right\}$ is the set of functions defined on nodes of the mesh, seven-point Laplacian stencil with suitable boundary condition (which will be discussed in section \ref{sec:4} is used for $\Delta f$ and $Kf(T)$ is defined by:

\begin{equation}
Kf(T)=\begin{cases}
\frac{||Df(T)||_F^{2}}{\det(Df(T))^{2/3}} & \text{if }\text{det}\left( Df(T) \right)>0,\\
\text{+\ensuremath{\infty}} & \text{otherwise}
\end{cases}
\end{equation}

\section{Algorithm}
\label{sec:4}
In this section, we explain the numerical algorithm to optimize the energy functional described in the last section. For the ease of the explanation, we will demonstrate the numerical algorithms for the 3-D case. The numerical algorithms for general $n$-D spaces can be done similarly. We split the optimization problem (\ref{discreteenergy}) as follows:

\begin{equation}\label{splittedorginal}
\inf_{f,R}\sum_{\text{tetrahedron }T}K(f,R,T)+\frac{\sigma}{2}\sum_{\text{node }x}\Vert \Delta f(x) \Vert_2^2 \quad\text{given }R(T)=Df(T)
\end{equation}

\noindent where:

\begin{equation}\label{splitenergy}
K(f,R,T)=\begin{cases}
\frac{||Df(T)||_F^{2}}{\det(R(T))^{2/3}} & \text{if }\text{det}\left( R(T) \right)>0,\\
\text{+\ensuremath{\infty}} & \text{otherwise}
\end{cases}
\end{equation}

We apply the alternating direction method of multipliers (ADMM) to optimize (\ref{splitenergy}).  ADMM was firstly proposed in 1975 and has received lots of attention recently due to its simple implementation and extensive applications to image processing and compressive sensing. We will briefly describe the idea of ADMM. For details, we refer the readers to \cite{ADMM1,ADMM2,ADMM3}. For a general optimization problem
\begin{equation}\label{generalADMM}
\inf_{x,y} E(x,y) \text{ subject to } y = Ax
\end{equation}
\noindent where $A\in M_{m\times n}(\mathbb{R})$. The augmented Lagrangian associated to the problem (\ref{generalADMM}) is given by
\begin{equation}\label{ADMM2}
L(x,y,\mathbf{\lambda},\mu) = E(x,y) + \frac{\mu}{2}||Ax-y -\lambda||^2 
\end{equation}

The alternating direction method with multiplier (ADMM) decouples the optimization process, which can be described as follows:
\begin{equation}
\begin{split}
x^{k+1} & = \mathbf{argmin}\{L(x,y^k,\mathbf{\lambda}^k,\mu^k)\}\\
y^{k+1} & = \mathbf{argmin}\{L(x^{k+1},y,\mathbf{\lambda}^k,\mu^k)\}\\
\mathbf{\lambda}^{k+1} &= \mathbf{\lambda}^k + \mu_k(Ax^{k+1}- y^{k+1})
\end{split}
\end{equation}
\noindent where $\{\lambda^k\}$ is the sequence approximating the Lagrange multiplier of the constraint $Ax = y$ and $\{\mu_k\}$ is a sequence of positive real number, called the {\it penalty parameters}. A variants of the choices of $\{\lambda^k\}$ and $\{\mu_k\}$ have been proposed. In other words, ADMM firstly solves for $x^{k+1}$ by fixing $y = y^k$, and then solves for $y^{k+1}$ by fixing $x = x^{k+1}$. This leads to efficient and parallelizable optimization algorithm.

Applying the ADMM to our problem, our numerical algorithm can now be described as follows. Suppose $(f^k,R^k,\lambda_k,\mu_k)$ is obtained at the $k$-th iteration. We first solve the f-subproblem:
\begin{equation}
\begin{split}
& f^{k+1}  = \underset{f}{\text{argmin}} \sum_{T}K(f,R^k,T)+\frac{\mu}{2}\sum_{T}||R^k-Df+\lambda^{k}||_F^{2} + \frac{\sigma}{2} \sum_{x} \Vert \Delta f \Vert_2^2 \\
&\text{ subject to } f^{k+1}(p_i) = q_i \text{ for } i = 1,..., m.
\end{split}
\end{equation}

\noindent We then solve the R-subproblem:
\begin{equation}
\begin{split}
& R^{k+1}(T) = \underset{\text{det}(R) > 0}{\text{argmin}}K(f^{k+1},R,T)+\frac{\mu}{2}||R-Df^{k+1}+\lambda^{k}||_F^{2} \text{ for each tetrahedron } T.
\end{split}
\end{equation}

The Lagrange multiplier $\lambda_{k+1}$ and the penalty parameter $\mu_{k+1}$ are updated as follows. 
\begin{align}
\lambda^{k+1}&= \lambda^{k}+R^{k+1}-Df^{k+1};\\
\mu_{k+1} &= \max \{\underset{T}{\max}\frac{30}{\det\left(R(T)\right)^{2/3}}, \mu_k\}.
\end{align}

The overall algorithm can now be summarized as Algorithm \ref{alg::LMalgorithm}.

\begin{algorithm}[h]
\KwIn{Domain $\Omega$; landmark sets $\{p_i \in \Omega \}_{i=1}^m$ and $\{q_i \in \Omega \}_{i=1}^m$.}
\KwOut{Landmark registration $f^*:S_1\to S_2$.}
\BlankLine
Initial $f^1 = \text{Identity map} \mathbb{I} $; $R^1 = Df^1$; $\lambda^1 = 0$\;
\Repeat{$||f^{k+1} - f^{k}||_{\infty} \leq \epsilon$}{
$f^{k+1} \leftarrow \underset{f}{\text{argmin}} \sum_{T}K(f,R^k,T)+\frac{\mu}{2}\sum_{T}||R^k-Df+\lambda^{k}||_F^{2} + \frac{\sigma}{2} \sum_{x} \Vert \Delta f \Vert_2^2 $\  \quad \quad \quad \quad \quad \quad \quad \quad \quad \quad subject to $f^{k+1}(p_i) = q_i $ for $i = 1,\ldots,m$\;
$R^{k+1} \leftarrow \underset{\text{det}(R) > 0}{\text{argmin}}K(f^{k+1},R,T)+\frac{\mu}{2}||R-Df^{k+1}+\lambda^{k}||_F^{2}$ for each tetrahedron $T$\;
$\lambda^{k+1} \leftarrow \lambda^{k}+R^{k+1}-Df^{k+1}$\;
Update $\mu_{k+1}= \max \{\underset{T}{\max}\frac{30}{\det\left(R(T)\right)^{2/3}}, \mu_k\}$\;
$k \leftarrow k + 1$\;
}
\caption{\it Quasi-conformal landmark-matching transformation algorithm \label{alg::LMalgorithm}}
\end{algorithm}

\bigskip

There are two subproblems in the algorithm, namely, the f-subproblem and the R-subproblem. In practice, we solve the $f$-subproblem first. In the following two subsections, we will explain how the f-subproblem and the R-subproblem can be solved in details.

\subsection{f-subproblem}
\label{sec:4_1}

The f-subproblem is to minimize the energy

\begin{equation}
E_{fsub}^k(f) =\sum_{T}\frac{||Df(T)||_F^{2}}{\det(R^k(T))^{2/3}}+\frac{\mu}{2}\sum_{T}||R^k(T)-Df(T)+\lambda^{k}(T)||_F^{2} + \frac{\sigma}{2} \sum_
{x} \Vert \Delta f \Vert_2^2 .\label{eq:f_org_org}
\end{equation}

Note that $f=(f_1,f_2,f_3):\Omega_1\subset \mathbb{R}^3\to \Omega_2\subset \mathbb{R}^3$ is a vector-valued function. The energy functional $E_{fsub}^k$ can be decoupled into $E_{fsub}(f) = E_{fsub1}^k(f_1) + E_{fsub2}^k(f_2) + E_{fsub3}^k(f_3)$. The optimization problem can be solved component-wisely. Therefore, we can regard $f$ as a scalar function only in this section. The corresponding Euler-Lagrange equation for this problem is of the form

\begin{equation}
\begin{cases}
\begin{array}{rcl}
\sigma \Delta^2 f(x) - \nabla\cdot(A(x)\nabla f(x)) & = & g(x);\\
f(p_{i}) & = & q_{i},
\end{array}\end{cases}\label{eq:f_org}
\end{equation}
where $A(x)$ is a diagonal matrix with diagonal entries

\begin{equation}
\sum_{\text{six } T \text{ touch the corresponding edge}} \left(\frac{1}{\text{det}(R(T))^{2/3}}+\mu \right),
\end{equation}

\noindent and $g(x) = -\mu \nabla \left(R^k(T) + \lambda^k(T) \right)$. Subtracting both side of \eqref{eq:f_org} by any function that satisfies $f(p_{i})=q_{i}$, we can assume $f(p_{i})=0$. 

Equation (\ref{eq:f_org}) can be discretized into a linear system. To solve equation (\ref{eq:f_org}), we apply the preconditioned conjugate gradient (PCG) method \cite{saad2003iterative}. In order to apply the PCG method, a suitable preconditioner approximating the inverse of the coefficient matrix of (\ref{eq:f_org}) must be chosen. In this work, we use the multi-grid V-cycle of an approximated linear system of (\ref{eq:f_org}) as the preconditioner $M$ \cite{Tatebe}. Equation (\ref{eq:f_org}) is then solved by PCG with the preconditioner matrix $M$. 

We will now explain how the multi-grid V-cycle preconditioner $M$ is constructed. Note that the penalty parameter $\mu$ in the ADMM aims to drive $R$ to be closer to $Df$, so that optimal solution eventually satisfies the constraint $R=Df$. If $\mu$ is too small, the solution in each ADMM iteration may be far away from the admissible solution satisfying the constraint $R=Df$. It may take a long time to converge to the optimizer of (\ref{splittedorginal}) satisfying the constraint $R=Df$. On the other hand, if $\mu$ is too big, the solution in each ADMM iteration better satisfy the constraint $R=Df$. But again, it may take a long time to obtain the optimizer minimizing the energy function of (\ref{splittedorginal}). Hence, an optimal penalty parameter has to be carefully chosen \cite{Ghadimi}. In our algorithm, the penalty parameter $\mu$ is chosen to be $\mu \geq \underset{T}{\max}\frac{30}{\det\left(R(T)\right)^{2/3}}$. This parameter is good enough for the ADMM converges at a reasonable rate.

With this parameter, we approximate $A(x)=6\mu \mathbf{I}$. Hence, equation (\ref{eq:f_org}) can be approximated by a Poisson equation

\begin{equation}
\begin{cases}
\begin{array}{rcl}
\sigma \Delta^2 f(x) - 6\mu\Delta f(x) & = & g(x),\\
f(p_{i}) & = & 0.
\end{array}\end{cases}\label{eq:f_new}
\end{equation}

We proceed to approximate the solution of the above approximated system to get a preconditioner $M$. We remark that the above approximated system is introduced to obtain the preconditioner $M$. With the preconditioner $M$, the original f-subproblem (\ref{eq:f_org}) will be solved exactly using the PCG method.

\noindent If $\sigma \neq 0$, the equation (\ref{eq:f_new}) can be split into two coupled Poisson equations

\begin{equation}
\begin{cases}
\begin{array}{rcl}
-h-\Delta f & = & 0,\\
-\sigma \Delta h- 6\mu \Delta f& = & g,\\
f(p_{i}) & = & 0.
\end{array}\end{cases}\label{eq:f_system1}
\end{equation}

\noindent If $\sigma = 0$, the equation (\ref{eq:f_new}) can be simplified to be the following Poisson equation
\begin{equation}
\begin{cases}
\begin{array}{rcl}
- 6\mu\Delta f(x) & = & g(x),\\
f(p_{i}) & = & 0.
\end{array}\end{cases}\label{eq:f_system2}
\end{equation}

\noindent In both cases, the equation can be approximately solved using the multi-grid V-cycle, which gives us a preconditioner $M$ for solving (\ref{eq:f_org}).

We will now explain the multi-grid V-cycle briefly. For details, we refer the readers to \cite{MultigridBook,briggs2000multigrid}. 

Let us first define a hierarchy of discretization of the unit cube, that is $V_{1}\subset V_{2^{-1}}\subset\cdots\subset V_{2^{-J}}$ where $V_{h}$ is a uniform grid on unit cube with spacing $h$. On $V_{h}$, we discretize the equation \eqref{eq:f_system1} and \eqref{eq:f_system2} respectively as

\begin{minipage}{.48\textwidth}
\begin{equation}
\begin{cases}
\begin{array}{rcl}
-\left( \begin{array}{cc} I & L_h \\ L_h & 6\frac{\mu}{\sigma} L_h \end{array}\right) \left( \begin{array}{c} h \\ f \end{array} \right) & = & \left( \begin{array}{c} 0 \\ g \end{array}\right),\\
f(p_{i}^{h}) & = & 0.
\end{array}\end{cases}\label{eq:f_system1_discerete}
\end{equation}
\end{minipage}
\begin{minipage}{.48\textwidth}

\begin{equation}
\begin{cases}
\begin{array}{rcl}
-6\mu L_{h}f(x) & = & g,\\
f(p_{i}^{h}) & = & 0.
\end{array}\end{cases}\label{eq:f_system2_discerete}
\end{equation}
\end{minipage}

\noindent where $L_{h}$ is the seven-point Laplacian stencil with suitable boundary conditions and $p_{i}^{h}$ is the landmark points on the grid $V_{h}$. To simplify, let $\mathcal{L}_h = \left( \begin{array}{cc} I & L_h \\ L_h & 6\frac{\mu}{\sigma} L_h \end{array}\right)$ if $\sigma \neq 0$ and $\mathcal{L}_h = 6\mu L_h$ if $\sigma = 0$.

The boundary conditions depends on the setting of the original problem \eqref{discreteenergy}. Either Dirchlet, Neumann or the combination of both can be enforced. For example, suppose the unit cube is mapped to a unit cube, the boundary conditions can be set as follows. Let $f=(f_{1},f_{2},f_{3})$. Then:
\begin{equation}
\begin{split}
f_{1}(0,y,z)&=0\text{ and }f_{1}(1,y,z)=1,\\
f_{2}(x,0,z)&=0\text{ and }f_{2}(x,1,z)=1,\\
f_{3}(x,y,0)&=0\text{ and }f_{3}(x,y,1)=1.
\end{split}
\end{equation}

Therefore, for $f_1$, we impose the Dirichlet boundary condition on $\{0,1\}\times[0,1]\times[0,1]$ and Neumann boundary condition on other boundaries. For $f_{2}$ and $f_{3}$, we do it similarly.

The next question is how we set the landmark constraints on coarser levels. In other words, we need to choose vertices $\{p_i\}_{i=1}^m$ such that $f(p_i) = 0$. On the finest grid, these points are chosen to be the original landmark points from the input. On the next coarser grid, these points are chosen to be the set of points belonging to the neighbourhood of landmark points at the previous finer level. For example, if we have a landmark point $(0.375,0.375,0.375)$ on $V_{2^{-2}}$, then on $V_{2^{-1}}$, the set of points in the neighbourhood of it are $(0.25,0.25,0.25)$, $(0.25,0.25,0.5)$, $(0.25,0.5,0.25)$, $(0.25,0.5,0.5)$ and so on. All these points will be selected as landmark points on $V_{2^{-1}}$. Although this scheme would probably make the coarsest level having many landmark points, it does not cause either convergence or complexity problem because more landmark points means less free variables and faster convergence.

The multi-grid V-cycle can now be described as follows. At level $h$ (the grid with spacing $h$), \eqref{eq:f_system1_discerete} and \eqref{eq:f_system2_discerete} can both be regarded as a linear system of the form $\mathcal{L}_h f = {\bf c}$. We first relax $\mathcal{L}_h f = {\bf c}$ using certain iterative scheme, such as Jacobi, Gauss-Seidel or Successive over-relaxation methods. We denote the approximated solution $\tilde{f}_h$ after the relaxation by $\tilde{f}_h = S(f,{\bf c})$. We then compute the residual $r_h = {\bf c}- \mathcal{L}_h \tilde{f}_h$. To improve the solution, we relax $\mathcal{L}_{2h} e = r_{2h}$ on a coarser grid $V_{2h}$, where $r_{2h} = I_h^{2h}(r_h)$ is the projection of $r_h$ from level $h$ to level $2h$ and $I_h^{2h}$ is the linear projection operator. Denote the approximated solution of $\mathcal{L}_{2h} e = r_{2h}$ by $e_{2h}$. Then, the approximated solution of $\mathcal{L}_h f = {\bf c}$ can be improved by $\tilde{f}_h$ by $\tilde{f}_h\leftarrow \tilde{f}_h +I_{2h}^{h} e_{2h}$. This completes a V-cycle at level $h$. Note that when computing the approximated solution of $\mathcal{L}_{2h} e = r_{2h}$, we can again apply a multi-grid V-cycle on level $2h$.

\begin{figure}[t]
        \centering
        \includegraphics[width=2.5in]{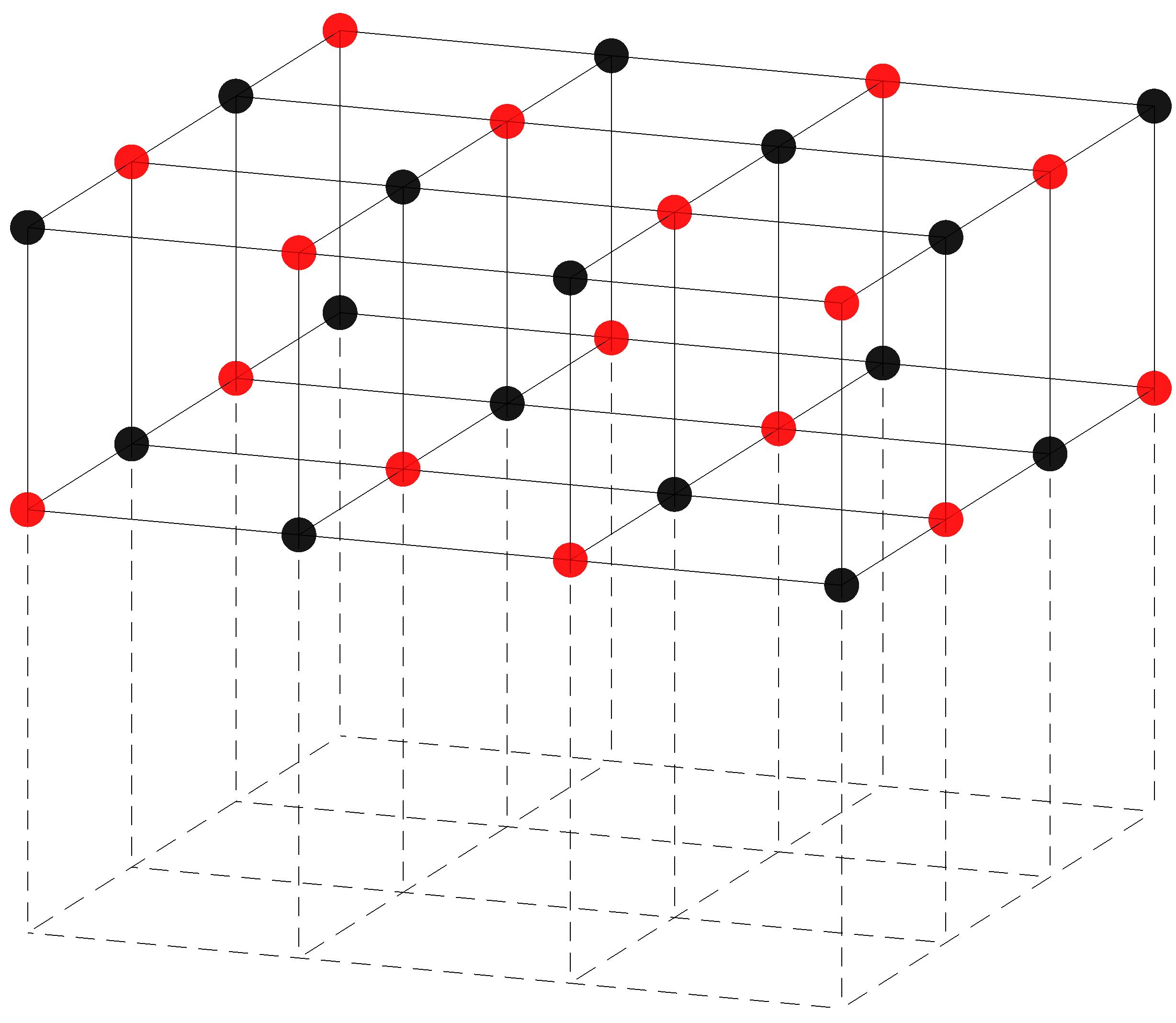}
\caption{Illustration of red-black ordering of grid points for 3-dimensional space. \label{alg::onestep_Vcycle}}
\end{figure}

 The detailed multi-grid V-cycle algorithm can now be described as in Algorithm \ref{alg::onestep_Vcycle}.

\begin{algorithm}[h]
\BlankLine
If $h = 2^{-k}$ for some $k \geq 1$, return $\mathcal{L}_{h}^{-1}g$\;
$f\leftarrow S(f,g)$\;
$f\leftarrow f+I_{2h}^{h}\text{Vcycle}_{2h}\left(0,I_{h}^{2h}(g-\mathcal{L}_{h}f)\right)$\;
$f\leftarrow S(f,g)$\;
\caption{\it $f = \text{Vcycle}_h(f,g)$}
\end{algorithm}

The relaxation $S$ in Algorithm \ref{alg::onestep_Vcycle} removes the high frequency component in the residual for $r_h = {\bf c } - \mathcal{L}_h \tilde{f}_h$. In this paper, the relaxation $S$ is chosen to be the Red-black Gauss-Seidel (RBGS) iterations. We will briefly describe the RBGS iteration. For details, we refer the readers to \cite{saad2003iterative}. The red-black Gauss-Seidel modifies the standard Gauss-Seidel method by reordering different equations. The basic idea is to group the grid points into two groups, identified as black and red nodes, such that black nodes are surrounded by red nodes only and vice versa. The red-black grouping of grid points in 3-dimensional space is as shown in Figure . The Laplace operator $\mathcal{L}_h$ 
under the red-black ordering of grid points can be rewritten as: $\mathcal{L}_h = \left(\begin{array}{cc}\mathbf{D}_r & \mathbf{U}\\
\mathbf{L} & \mathbf{D}_b\end{array}\right)$, where $\mathbf{D}_r$ and $\mathbf{D}_b$ are diagonal matrices associated to the red nodes and black nodes respectively. The Gauss-Seidel iteration can now be written as:
\begin{equation}
\begin{split}
\tilde{f}_h^{r,n+1} = \mathbf{D}_r^{-1} (-\mathbf{U}\tilde{f}_h^{b,n} + \mathbf{c}_r),\\
\tilde{f}_h^{b,n+1} = \mathbf{D}_b^{-1} (-\mathbf{L}\tilde{f}_h^{r,n+1} + \mathbf{c}_b)
\end{split}
\end{equation}
\noindent where $\tilde{f}_h^{r,n}$ and $\tilde{f}_h^{b,n}$ are the components of $\tilde{f}_h$ associated to the red and black nodes respectively at the $n$-th iterations. $\mathbf{c}_r$ and $\mathbf{c}_b$ are the components of $\tilde{c}$ associated to the red and black nodes respectively. As a result, instead of solving a triangular system as in the standard Gauss-Seidel iterations, we perform matrix-vector products and vector scaling operations with half as many variables in each iteration.

In our implementation, we have chosen $S$ to be four iterations of Red-Black Gauss-Seidel method. The red-black ordering for step 2 in Algorithm \ref{alg::onestep_Vcycle} is the opposite to the red-black ordering for step 4. This reverse ordering ensure the obtained multi-grid V-cycle preconditioner $M$ to be symmetric positive definite for applying PCG \cite{Tatebe}.

For the restriction and interpolation operator, the full weighting restriction and bilinear interpolation operator are used. Note that the result of the interpolation operator satisfies the landmark points condition because of our choices of landmark points on the coarse grid.

The overall $\text{Vcycle}_{h}(0,g)$ is a linear operator on $\tilde{f}$. We simply write it as $M\tilde{f}$. $M$ is our desired preconditioner. With $M$, we apply the PCG method with the preconditioner matrix $M$ to solve the original f-subproblem \eqref{eq:f_org} \cite{MultigridBook}. This can be described as follows.

\begin{algorithm}[h]
\label{alg::f_problem_minimize}
\BlankLine
Denote $Mg=\text{Vcycle}_{h}(0,g)$\;
Apply the preconditioned conjugate gradient method on \eqref{eq:f_org} with the preconditioner matrix $M$\;
\caption{\it Solving \eqref{eq:f_org}}
\end{algorithm}

\subsection{R-subproblem}
\label{sec:4_2}
The R-subproblem in Algorithm \ref{alg::LMalgorithm} is a tetrahedron-wise problem. Therefore, parallel computing can be adopted in this subproblem. More explicitly, we want to find $R(T)$ on each tetrahedron $T$ which minimizes the following energy
\begin{equation}\label{eq:Rproblem}
\underset{R(T) \in \mathbb{R}^{3 \times 3},\text{det}(R(T)) > 0}{\text{min}}\left(\frac{||Df^k||_F^{2}}{\det(R(T))^{2/3}}+\frac{\mu}{2}\Vert R(T)- B \Vert_F^{2}\right)
\end{equation}
where $B=Df^k-\lambda^{k}$. 

Let the SVD of $B$ be $U\Sigma V^{*}$. In the case when $R$ can be written as $U\tilde{\Sigma} V^{*}$, where $\tilde{\Sigma}$ is an unknown diagonal matrix. Problem (\ref{eq:Rproblem}) can be much simplified. Since we want $\text{det}(R(T))>0$, $\tilde{\Sigma}$ must satisfy $\text{sgn}(\text{det}(\tilde{\Sigma})) = \text{sgn}(\text{det}(UV^*))$. Denote $\Gamma = \{ \tilde{\Sigma} \ | \ \text{sgn}(\text{det}(\tilde{\Sigma})) = \text{sgn}(\text{det}(UV^*)) \}$. By substituting these representations and constraint to problem (\ref{eq:Rproblem}), we have the following optimization problem with three variables:
\begin{equation}
\underset{\tilde{\Sigma}\in \Gamma}{\text{min}}\left( \frac{||Df^k||^{2}}{\det(\tilde{\Sigma})^{2/3}}+\frac{\mu}{2}\Vert \tilde{\Sigma} - \Sigma \Vert_F^{2} \right)
\label{eq:simplify_Rprob}
\end{equation}

In other words, by making the assumption that $R$ can be written as $U\tilde{\Sigma} V^{*}$, we can simply the original R-subprobem as \eqref{eq:simplify_Rprob}. A natural question is the relationship between the problem  \eqref{eq:simplify_Rprob} and our original R-subproblem \eqref{eq:Rproblem}. The following theorem gives the answer.

\begin{theorem}
Suppose $\overline{\Sigma}$ is the minimizer of (\ref{eq:simplify_Rprob}). Then, $R = U\overline{\Sigma}V^*$ is the minimizer of (\ref{eq:Rproblem}), where $B = U\Sigma V^*$ is the SVD of $B$.
\end{theorem}
\begin{proof}
This is related to the general two-sided Procrustes problem. Suppose $X_1$ and $X_2$ are $n\times n$ matrices. Define:
\begin{equation}
E_P(Q_1,Q_2) = ||Q_1^* X_1 Q_2 - X_2||_F^2
\end{equation}
\noindent where $Q_1$ and $Q_2$ are $n\times n$ orthogonal matrices. Let $X_1 = P_1 \Sigma_1 R_1^*$ and $X_2 = P_2 \Sigma_2 R_2^*$ be the SVDs of $X_1$ and $X_2$ respectively. Then, the minimizer of $E_P$ satisfies:
\begin{equation}\label{eq:procrustecondition}
P_1 = Q_1 P_2 \Pi; \ \ \ R_1 = Q_2 R_2 \Pi,
\end{equation}
\noindent where $\Pi$ is the permutation matrix that maximizes $\mathrm{Tr}(\Sigma_2^* \Pi^* \Sigma_1 \Pi)$ (see p.89-90 in \cite{gower2004procrustes}).

Let $B = U\Sigma V^*$ be the SVD of $B$ and let $P\widetilde{\Sigma}Q$ be the SVD of $U^*RV$. Our R-subproblem (\ref{eq:Rproblem}) is equivalent to minimizing:
\begin{equation}\label{eq:ERnew}
E_R^{new}(\widetilde{P},\widetilde{\Sigma},\widetilde{Q}):=\left\{||\widetilde{P}\widetilde{\Sigma}\widetilde{Q}^*-\Sigma||_F^2 + \frac{c}{\mathrm{det}(\widetilde{\Sigma})^{2/3}}\right\},
\end{equation}
\noindent for some positive constant c.

Let $\overline{\Sigma}$ be the minimizer of:
\begin{equation}
\min_{\overline{\Sigma}}\left\{||\overline{\Sigma} - \Sigma||_F^2 + \frac{c}{\mathrm{det}(\widetilde{\Sigma})^{2/3}}\right\}.
\end{equation}

Fixing a diagonal matrix $D$, we consider the minimization problem over $(\widetilde{P},\widetilde{Q})$ of $E_R^{new}(\widetilde{P},D,\widetilde{Q})$. According to (\ref{eq:procrustecondition}), the minimizer must satisfy $I = \widetilde{P}\Pi$ and $I = \widetilde{Q}\Pi$. Thus, for any orthogonal matrices $P$ and $Q$ and diagonal matrix $D$,
\begin{equation}
\begin{split}
E_P^{new} (P,D,Q) &= ||PDQ^*-\Sigma||_F^2 + \frac{c}{(\mathrm{det}(D))^{2/3}}\\
&\geq ||\Pi^* D\Pi - \Sigma||_F^2 + \frac{c}{(\mathrm{det}(\Pi^*D\Pi))^{2/3}}\\
&\geq ||\overline{\Sigma} - \Sigma||_F^2 + \frac{c}{(\mathrm{det}(\overline{\Sigma})^{2/3}}\\
& = E_R^{new}(I,\overline{\Sigma},I).
\end{split}
\end{equation}

Thus, $(I,\overline{\Sigma},I)$ is a minimizer of (\ref{eq:ERnew}). We conclude that: $R = UI\overline{\Sigma}IV^* = U\overline{\Sigma}V^*$ is a minimizer of (\ref{eq:Rproblem}).
\end{proof}

\bigskip

\begin{theorem}
The Euler Lagrange equation of (\ref{eq:simplify_Rprob}) is:
\begin{equation}
\tilde{\Sigma}-\frac{a}{(\text{det}(\tilde{\Sigma}))^{2/3}}\tilde{\Sigma}^{-1}=\Sigma, \quad \text{where } a=\frac{2 \Vert Df^k \Vert ^{2}}{3\mu}.
\label{eq:R_prob_EL}
\end{equation}
\end{theorem}
\begin{proof}
Let $x_{i}$ be the diagonal of $\Sigma$ and $y_{i}$ be the diagonal of $\tilde{\Sigma}$. Denote $\tilde{y}_1 = y_1 + \epsilon \lambda_1$ to be the variation of $y_1$. Consider the derivative of energy with respective to $\epsilon$, we have
\begin{equation}
\begin{split}
 \frac{d}{d \epsilon} \frac{\Vert Df^k \Vert^2}{\left(\tilde{y}_1y_2y_3\right)^{2/3}} + \frac{\mu}{2}\left[(\tilde{y}_1-x_1)^2 + (y_2-x_2)^2 + (y_3-x_3)^2\right] \biggr\rvert_{\epsilon = 0} & = 0 \\
\Rightarrow  \lambda_1 \left( \frac{-a}{\left(y_1y_2y_3\right)^{2/3}}\left(\frac{1}{y_1}\right) + y_1-x_1\right) & = 0.
\end{split}
\end{equation}
\noindent Since $\lambda_1$ is arbitrary, we have
\begin{equation}
\frac{-a}{\left(y_1y_2y_3\right)^{2/3}}\left(\frac{1}{y_1}\right) + y_1-x_1 = 0
\end{equation}
\noindent Similar equations can be obtained for the variations of $y_2$ and $y_3$. By combining the results, we have the same formula as in (\ref{eq:R_prob_EL}).
\end{proof}

To tackle with the nonlinear recurrence equation (\ref{eq:R_prob_EL}), we propose Algorithm \ref{alg::R_prob_solve} below that gives the solution of (\ref{eq:recur_eqn}) to obtain a minimizer of the optimization problem (\ref{eq:simplify_Rprob}). More specifically, since the system (\ref{eq:R_prob_EL}) is coupled by the term $\det(\tilde{\Sigma})$, we can solve the equation
iteratively by
\begin{equation}
\tilde{\Sigma}_{n}-\frac{a}{\left(\text{det}(\tilde{\Sigma}_{n-1})\right)^{2/3}}\tilde{\Sigma}_{n}^{-1}=\Sigma
\label{eq:recur_eqn}
\end{equation}
where $\tilde{\Sigma}_{n}$ is the $\tilde{\Sigma}$ in step $n$. Define $D_n = \text{det}(\tilde{\Sigma}_{n-1})^{2/3}$. By element-wise decoupling the nonlinear recurrence equation (\ref{eq:recur_eqn}), we have the quadratic equations $y_i^n - \frac{a}{D}(y_i^n)^{-1} = x_i$ for $i = 1,2,3$, where $\tilde{\Sigma}_{n} = \text{diag}(y_1^n,y_2^n, y_3^n)$. Solving the quadratic equations, we have
\begin{equation}
y_{i}^n(D_n)=\frac{x_{i}\pm\sqrt{x_{i}^{2}+\frac{4a}{D_n}}}{2},\quad i = 1,2,3
\label{eq:y_eq}
\end{equation}
where the sign is chosen according to Algorithm \ref{alg::R_prob_solve}. The motivation and the convergence analysis of the proposed iteration scheme is explained in Theorem 3.
\begin{algorithm}[h]
\BlankLine
Compute the SVD of $B=U\Sigma V^{*}$ where the diagonal of $\Sigma$ is $x_{i} \geq 0 $\;
Set $D^1 = (\text{det}(R^{(\text{last})})^{2/3}$. Denote $D_{n+1} = (y^{n}_1y^{n}_2y^{n}_3)^{2/3}$\;
\Repeat{$||D_{n+1}-D_{n}||_{\infty}<\varepsilon$}{
Set $y_{i}^{n} = \frac{1}{2}\left(x_{i}+\sqrt{x_{i}^{2}+\frac{4a}{D^{n}}}\right)$ for $i=1,2,3$\;
If $\text{det} (UV^*) < 0$, set $y_{i}^{n} = \frac{1}{2}\left(x_{i}-\sqrt{x_{i}^{2}+\frac{4a}{D_{n}}}\right)$ for $i=\text{argmin}_{i}x_{i}$\;
$D^{n+1} \leftarrow \frac{1}{2}\left(D_{n}+\left(y_{1}^{n}y_{2}^{n}y_{3}^{n}\right)^{2/3}\right)$\;
$n \leftarrow n + 1$\;
}
$R = U\tilde{\Sigma}V^{*}$ where the diagonal entries of $\tilde{\Sigma}$ are $y_{i}$\;
\caption{\it Solving (\ref{eq:recur_eqn}) \label{alg::R_prob_solve}}
\end{algorithm}

Before introducing Theorem 3, the following lemma is necessary.

\begin{lemma}
Let $\sigma \in \mathbb{R}^n_{+}$ be a vector with positive values. The function
\begin{equation}
f(\sigma) = \frac{c}{\prod_{i=1}^n \sigma_i^{2/n}} + \mu \sum^n_{i=1} (\sigma_i - a_i)^2,\quad c>0 \text{ and } \mu > 0
\end{equation}
is convex.
\end{lemma}

\begin{proof}

Recall that the log barrier function $B(u) = -\log(\det(u))$, where $u$ is a symmetric positive definite matrix, is convex \cite{borwein2010convex}. This implies that $\frac{2}{n}B(D)$, where $D$ is a diagonal matrix with positive diagonal elements $\left\{d_{ii}\right\}_{i=1}^n$, is also a convex in $D$. Note that
\begin{equation}
\exp\left(-\frac{2}{n}B(D)\right) = \exp\left( -\frac{2}{n}\log(\det(u))\right) = \exp\left( \log\left( \prod_{i=1}^n \frac{1}{d_{ii}^{2/n}} \right) \right) = \frac{1}{\prod_{i=1}^n d_{ii}^{2/n}}.
\end{equation}
As the exponential of a convex function is also convex and $c > 0$, we have shown that $\frac{c}{\prod_{ii=1}^n \sigma_i^{2/n}}$ is convex.

Define $g(\sigma) = \sum^n_{i=1} (\sigma_i - a_i)^2$. We have
\begin{equation}
\frac{\partial^2 g}{\partial \sigma_j \sigma_i} = \left\{ \begin{array}{rl} 2 & \text{ if } i = j, \\ 0 & \text{ otherwise } \end{array} \right.
\end{equation}
Therefore, the Hessian matrix of $g$ is equal to $2I_n$, where $I_n$ is the n-dimensional identity matrix. Therefore $g$ is also convex in $\sigma$. By combining both results, we can conclude that $f(\sigma)$ is a convex function.
\end{proof}

The above lemma states that the simplified optimization problem \eqref{eq:simplify_Rprob} is convex in the positive octant region. In fact, using the same argument, we can show that the optimization problem \eqref{eq:simplify_Rprob} is convex in any one of the octant regions. Hence, the optimization is a global minimizer in each octant region.

Now, we will explain the convergence of Algorithm \ref{alg::R_prob_solve} to the minimizer of the optimization problem (\ref{eq:simplify_Rprob}).

\begin{theorem}
Given any $3\times3$ matrix $B$, $a>0$. Algorithm \ref{alg::R_prob_solve} converges linearly to a solution of the nonlinear recurrence equation (\ref{eq:R_prob_EL}) with the rate $\frac{1}{2}$, which is a minimizer of (\ref{eq:simplify_Rprob}).
\end{theorem}
\begin{proof}
Recall that $y_i^n$ at the $n^{\text{th}}$ iteration is defined as follows:
\begin{equation}
y_{i}^k(D_n)=\frac{x_{i}\pm\sqrt{x_{i}^{2}+\frac{4a}{D_n}}}{2}.
\end{equation}
The sign of $y_i^n$ is chosen as to minimize the energy functional (\ref{eq:simplify_Rprob}), which is given by the following:
\begin{equation}
\begin{split}
\underset{\tilde{\Sigma}\in \Gamma}{\text{min}}\left( \frac{||Df^k||^{2}}{\det(\tilde{\Sigma})^{2/3}}+\frac{\mu}{2}\Vert \tilde{\Sigma} - \Sigma \Vert_F^{2} \right) &= \mu \underset{\tilde{\Sigma}\in \Gamma}{\text{min}} \left(\frac{a}{\frac{2}{3}\det(\tilde{\Sigma})^{2/3}}+\frac{1}{2}\Vert \tilde{\Sigma} - \Sigma \Vert_F^{2}\right) \\
&=\mu \underset{\tilde{\Sigma}\in \Gamma}{\text{min}}\left( \frac{3a}{2(y_{1}y_{2}y_{3})^{2/3}}+\frac{1}{2}\sum\left(x_{i}-y_{i}\right)^{2} \right).
\end{split}
\end{equation}

Our goal is to make $D=(y_1y_2y_3)^{2/3}$ larger and $y_{i}$ closer to $x_{i}$. Therefore, the sign appears in equation (\ref{eq:y_eq}) can be determined according to the magnitude of the energy. If $\det (UV^*) > 0$, we can either set $+$ sign in equation (\ref{eq:y_eq}) for all $i$ or we set $-$ sign for only two of $y_i$. However, we can eliminate the second case by the following argument. Note that the second term in the energy functional dominates the overall energy and $x_i \geq 0 $ for all $i$, we have
\begin{equation}
\left(x_i - \frac{x_i + \sqrt{x_i^2 + \frac{4a}{D}}}{2}\right)^2 \leq \left(x_i - \frac{x_i - \sqrt{x_i^2 + \frac{4a}{D}}}{2}\right)^2 \quad \forall i.
\end{equation}
Therefore, the minimizer should satisfy the $+$ sign in equation (\ref{eq:y_eq}) for all $i$. If $\det (UV^*)<0$, we can either set $-$ sign in equation (\ref{eq:y_eq}) for all $i$ or we set $-$ sign for one of the $y_i$. Similar argument can be made and the minimizer should satisfy the $+$ sign in equation (\ref{eq:y_eq}) for all $i\neq\text{argmin}_{i}x_{i}$. This explains the purpose of step 5. Without loss of generality, we assume $\text{argmin}_{i}x_{i}=1$.

Let $F(D)=(y_{1}(D)y_{2}(D)y_{3}(D))^{2/3}$ and $G(D)=\frac{D+F(D)}{2}$.
We have
\begin{equation}
F'(D) = \frac{-2a}{3D^{2}}F(D)\sum_{i=1}^{3}\frac{\text{sgn}(y_i)}{y_{i}}\frac{1}{\sqrt{x_{i}^{2}+\frac{4a}{D}}}
\end{equation}
Note that each term $\frac{\text{sgn}(y_i)}{y_{i}}\frac{1}{\sqrt{x_{i}^{2}+\frac{4a}{D}}}$ in the sum is positive. Hence $F'(D)<0$.

\noindent For the case $\det (UV^*)>0$, we have
\begin{equation}
-F'(D) = \frac{2a}{3D^{2}}F(D)\sum_{i=1}^{3}\frac{1}{y_{i}}\frac{1}{\sqrt{x_{i}^{2}+\frac{4a}{D}}} < \frac{2a}{3D^{2}}F(D)\sum_{i=1}^{3}\frac{D}{2a} = \frac{F(D)}{D}.
\end{equation}

\noindent For the case $\det (UV^*)<0$, we have
\begin{equation}
\frac{\text{sgn}(y_1)}{y_{1}\sqrt{x_{1}^{2}+\frac{4a}{D}}} = \frac{2}{\sqrt{x_{1}^{2}+\frac{4a}{D}}-x_{1}}\frac{1}{\sqrt{x_{1}^{2}+\frac{4a}{D}}} = \frac{D}{2a}\frac{x_{1}+\sqrt{x_{1}^{2}+\frac{4a}{D}}}{\sqrt{x_{1}^{2}+\frac{4a}{D}}}
\end{equation}
when $i = 1$. For $i=2, 3$, we have
\begin{equation}
\frac{1}{y_{i}\sqrt{x_{i}^{2}+\frac{4a}{D}}} = \frac{2}{x_{i}^{2}+\frac{4a}{D}}\frac{\sqrt{x_{i}^{2}+\frac{4a}{D}}}{x_{i}+\sqrt{x_{i}^{2}+\frac{4a}{D}}} \leq \frac{D}{2a}\frac{\sqrt{x_{1}^{2}+\frac{4a}{D}}}{x_{1}+\sqrt{x_{1}^{2}+\frac{4a}{D}}}.
\end{equation}
Hence we have
\begin{align}
-F'(D) & \leq \frac{2a}{3D^{2}}F(D)\frac{D}{2a}\left(\frac{x_{1}+\sqrt{x_{1}^{2}+\frac{4a}{D}}}{\sqrt{x_{1}^{2}+\frac{4a}{D}}}+2\frac{\sqrt{x_{1}^{2}+\frac{4a}{D}}}{x_{1}+\sqrt{x_{1}^{2}+\frac{4a}{D}}}\right).
\end{align}
Since $\frac{1}{2}<\frac{\sqrt{x_{1}^{2}+\frac{4a}{D}}}{x_{1}+\sqrt{x_{1}^{2}+\frac{4a}{D}}}<1$, by considering the function $\frac{1}{x} + 2x$ on the interval $\left(\frac{1}{2},1\right)$, we conclude that the last term inside the parenthesis is less than $3$. Hence, in both cases, we have $-F'(D)\leq\frac{F(D)}{D}.$

Now, let $\tilde{D}$ be the solution of $G(D)=D$. We proceed to show that $\{D_n\}_{n=1}^{\infty}$ converges to $\tilde{D}$.

Consider the case when $D_n\geq\tilde{D}$. Since $G'(D) = \frac{1}{2} + \frac{F'(D)}{2}$, we have
\begin{equation}
\frac{1}{2} \geq G'(D) \geq \frac{1}{2}-\frac{F(D)}{2D} \geq \frac{1}{2}-\frac{F(\tilde{D})}{2D} = \frac{1}{2}-\frac{\tilde{D}}{2D} \geq 0.
\end{equation}
\noindent This suggests $G$ is an increasing function in $D$. Also, $\tilde{D} = G(\tilde{D}) = \frac{\tilde{D}+ F(\tilde{D})}{2}$ implies $\tilde{D} = F(\tilde{D})$. Hence,
$\frac{D+F(\tilde{D})}{2} = \frac{D+\tilde{D}}{2} \geq \frac{D+F(D)}{2} = G(D) \geq \tilde{D}$ for $D\geq \tilde{D}$. We get that
\begin{equation}
D_n\geq \frac{D_n+\tilde{D}}{2} \geq G(D_n) = D_{n+1} \geq \tilde{D}.
\end{equation}

\noindent $\{D_n\}_{n=1}^{\infty}$ is thus a decreasing sequence converging to some $D^*\geq \tilde{D}$. Also, from the previous inequalities, we observe that $\frac{D^* + \tilde{D}}{2}\geq D^*\geq \tilde{D}$, which gives $\tilde{D}\geq D^* \geq \tilde{D}$.
We conclude that $D^* = \tilde{D}$.

For the case $D_n\leq\tilde{D}$, we have $G(D)=\frac{D+F(D)}{2} \geq \frac{D+\tilde{D}}{2}$. Thus,
\begin{equation}
G(D_n)= D_{n+1} \geq \frac{D_n+\tilde{D}}{2}\geq D_n.
\end{equation}
\noindent Suppose $D_k \leq \tilde{D}$ for all $k>n$. Using a similar argument as before, we conclude that $\{D_n\}_{n=1}^{\infty}$ is an increasing sequence converging to $\tilde{D}$. Suppose $D_k > \tilde{D}$ for some $k>n$. From the previous conclusion, we can also get the same convergence result. That is, $D_n\to \tilde{D}$.

In both cases, we can show that $\frac{|D_{n+1}-D_{n}|}{|D_{n}-D_{n-1}|}\leq \frac{1}{2}$. As a result, the sequence $\{D_n\}_{n=1}^{\infty}$ converges at a rate $\frac{1}{2}$. Also, since $\tilde{\Sigma}_n = \text{diag}(y_1^n(D_n), y_2^n(D_n), y_3^n(D_n))$ depends on $D_n$, $\tilde{\Sigma}_n$ converges to a solution of the nonlinear recurrence equation (\ref{eq:R_prob_EL}) with a rate $\frac{1}{2}$.

In other words, algorithm \ref{alg::R_prob_solve} converges to a solution of the Euler-Lagrange equation of (\ref{eq:simplify_Rprob}). Depending on the sign of $\det(UV^*)$, our algorithm search for a critical point at the first octant region if $\det(UV^*)>0$ and at the other octant (with $x_i<0$) if $\det(UV^*)<0$. According to Lemma 1, the critical point must be the global minimizer of (\ref{eq:simplify_Rprob}) in the corresponding octant region. Hence, algorithm 4 converges to the minimizer of (\ref{eq:simplify_Rprob}).
\end{proof}

\section{Experimental Result}
\label{sec:6}
To validate the effectiveness of our proposed algorithm, experiments on synthetic examples have been carried out to compute 3-dimensional quasi-conformal landmark-matching transformation. We have also applied our proposed algorithm on lung CT images with respiratory deformations. Experimental results are reported in this section.
\begin{figure}[!h]
        \centering
        \begin{subfigure}[b]{0.35\textwidth}
                \includegraphics[width=\textwidth]{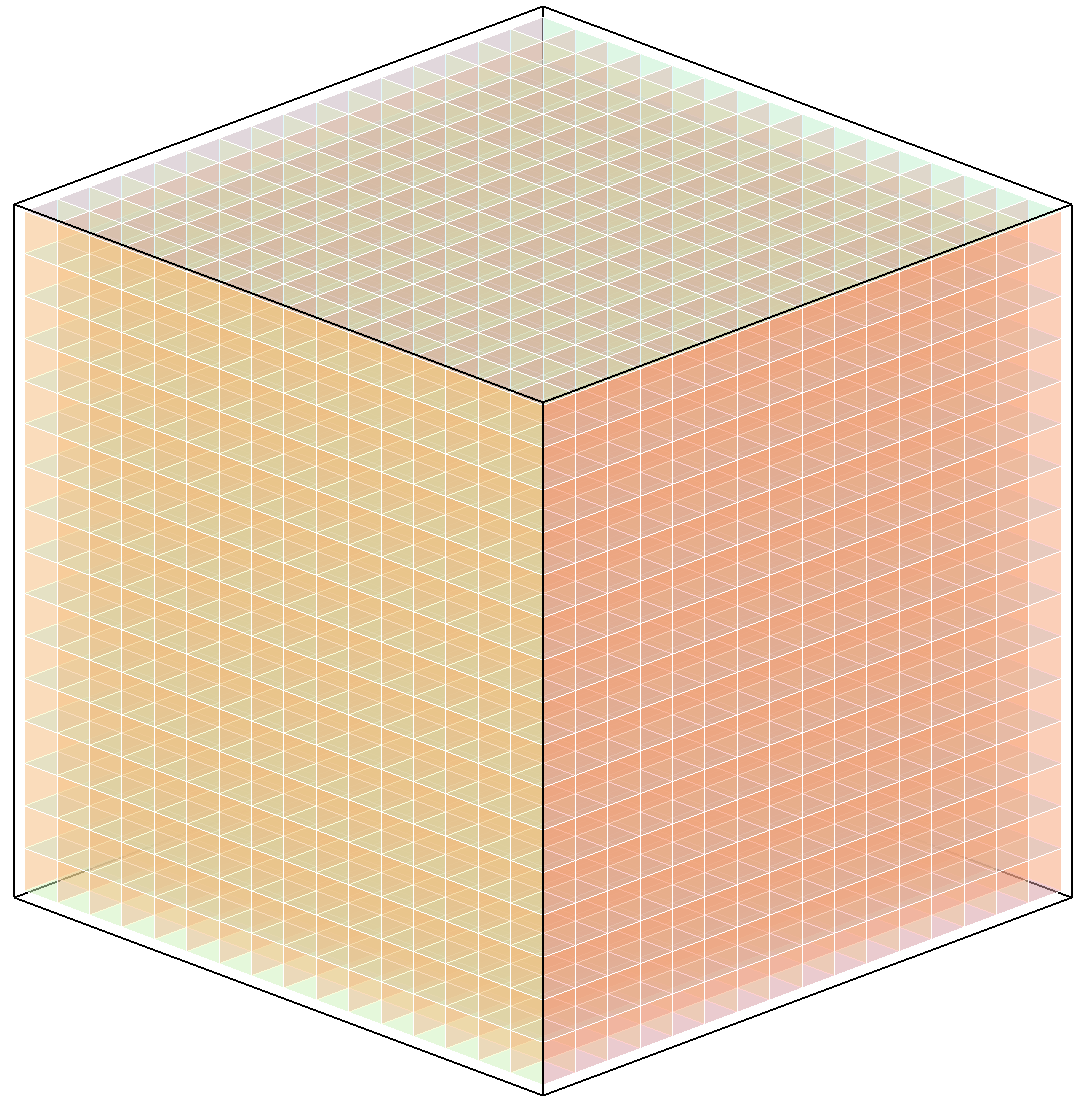}
	      \caption{Regular mesh on a cube.}
                \label{fig:reference_cube}
        \end{subfigure}
        \begin{subfigure}[b]{0.5\textwidth}
                \includegraphics[width=\textwidth]{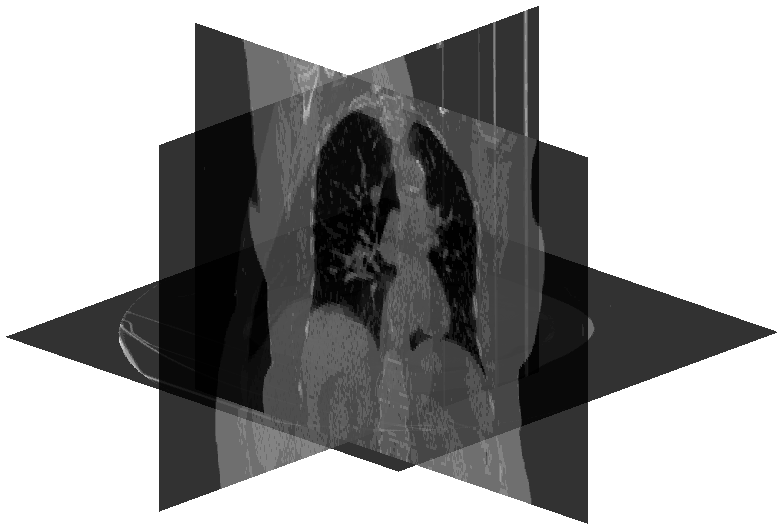}
	      \caption{CT image of a lung.}
                \label{fig:Lung3D_demo}
        \end{subfigure}
\caption{The regular reference mesh of the cube and the 3 dimensional lung CT image.}
\end{figure}
\subsection*{Synthetic examples}
\begin{figure}[t]
        \centering
        \begin{subfigure}[b]{0.33\textwidth}
                \includegraphics[width=\textwidth]{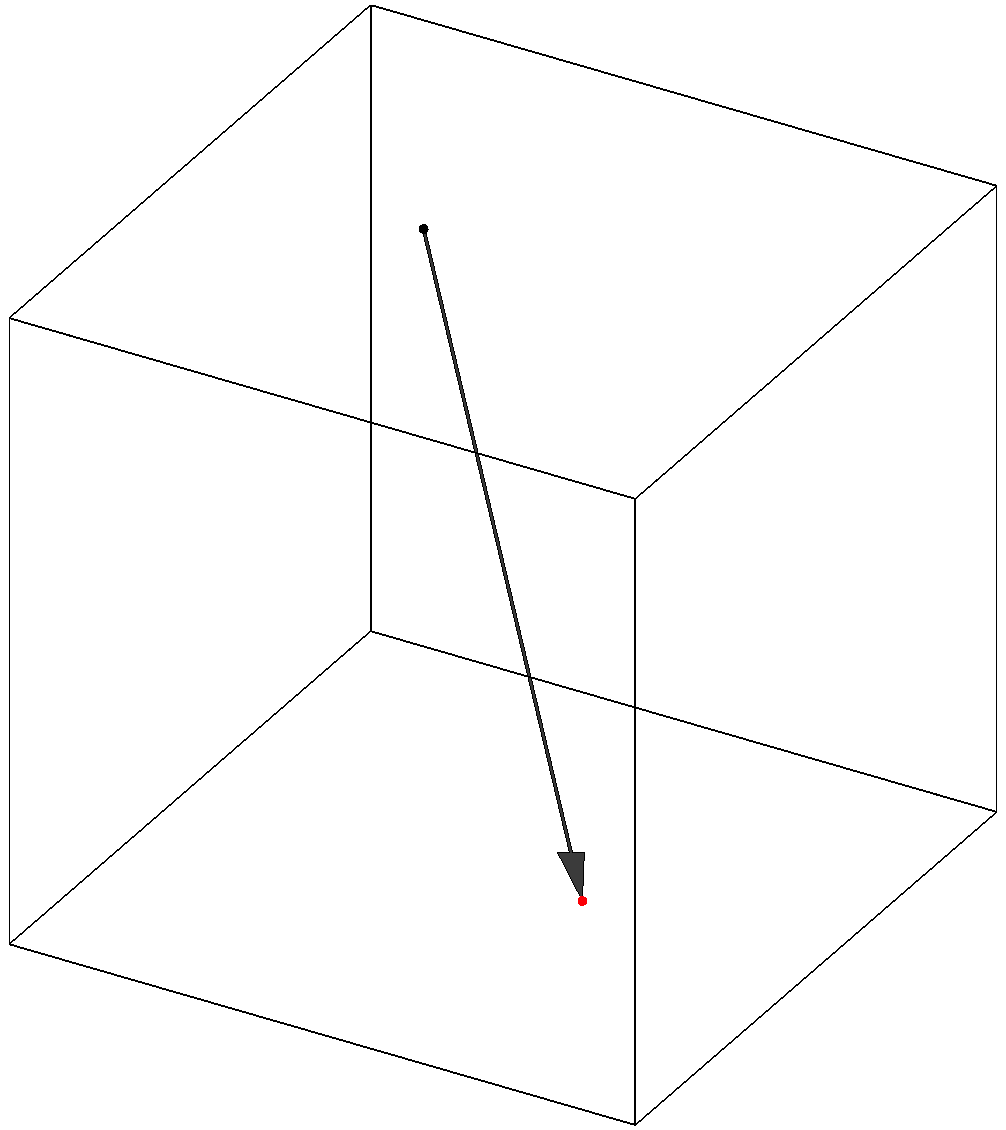}
	      \caption{Landmark}
                \label{fig:one_pointA}
        \end{subfigure}
        \begin{subfigure}[b]{0.32\textwidth}
                \includegraphics[width=\textwidth]{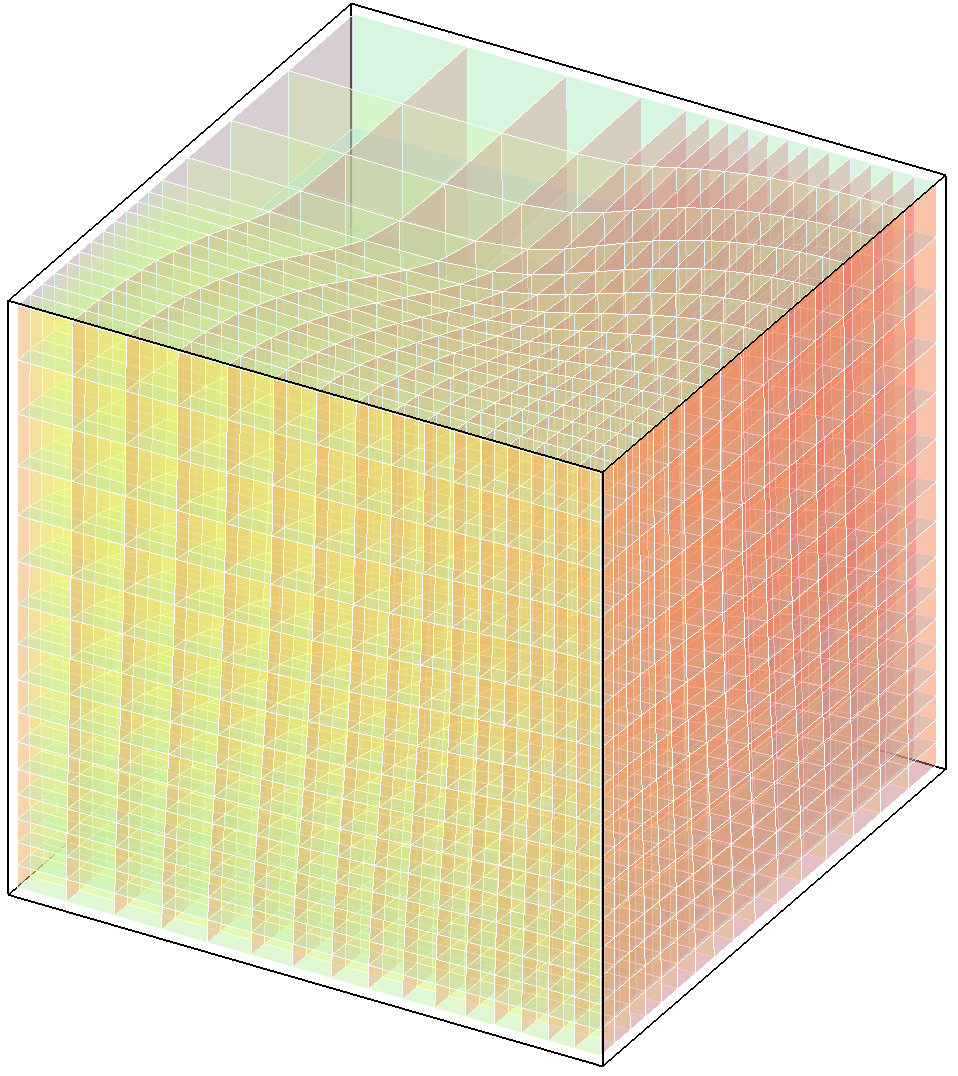}
	      \caption{QC deformation}
                \label{fig:one_pointB}
        \end{subfigure}
        \begin{subfigure}[b]{0.32\textwidth}
                \includegraphics[width=\textwidth]{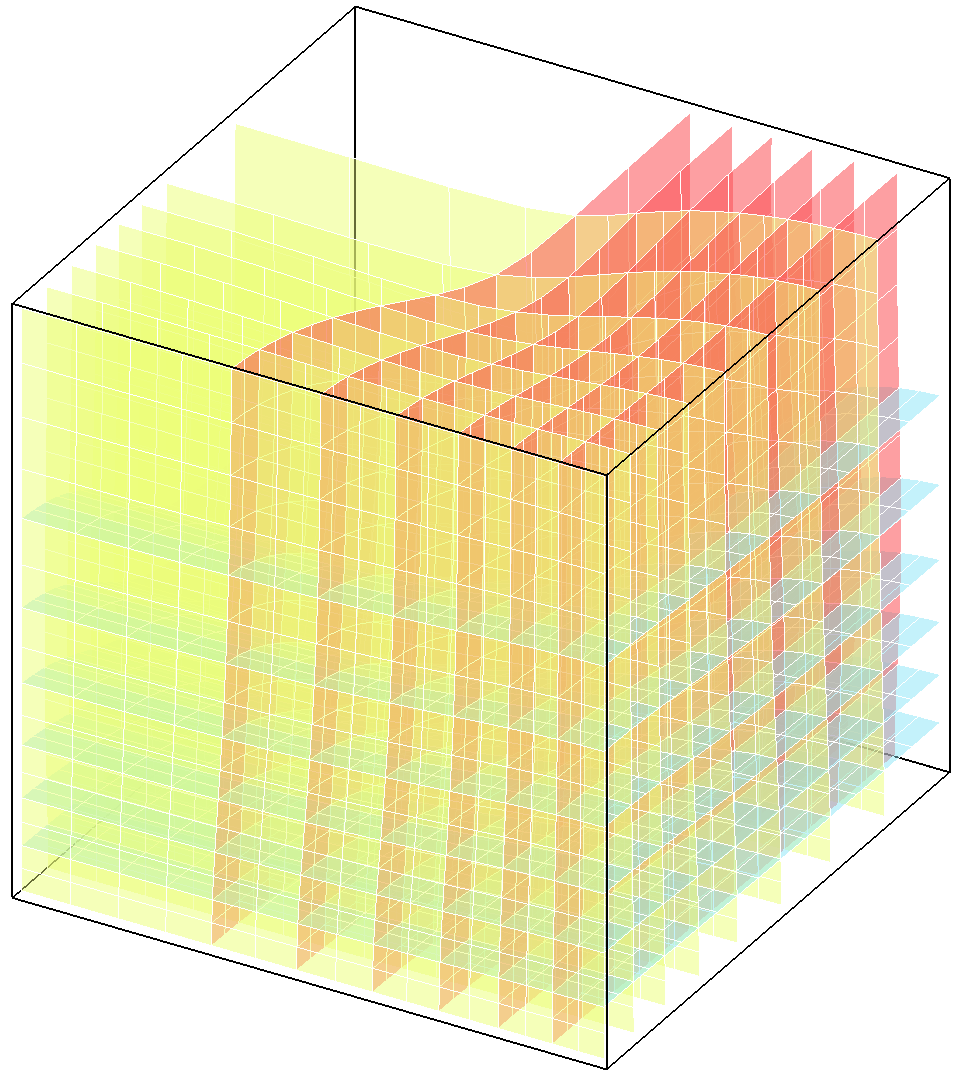}
	      \caption{Sparse view}
                \label{fig:one_pointC}
        \end{subfigure}
\caption{One-point landmark-matching experiment}
\end{figure}

We first test our algorithm to compute the landmark-matching transformation with one landmark. Figure \ref{fig:one_pointA} shows how the landmark point is deformed. The deformation of the landmark point is large. The point $p_1=[0.6,0.6,0.6]$ is moved to $q_1 = f(p_1)=[0.3,0.3,0.3]$. Using the proposed algorithm, we obtain a diffeomorphic transformation that satisfies the landmark constraint exactly. Figure \ref{fig:one_pointB} shows the obtained transformation. It is visualized by the deformation of the original reference mesh as shown in Figure \ref{fig:reference_cube} under the obtained transformation. The reference mesh is a regular grid of a cube discretizing the source domain. Figure \ref{fig:one_pointC} shows the visualization of the obtained transformation with a sparser view (to better demonstrate the transformation). Note that we set $\Omega_1 = \Omega_2 = \Omega = [0,1]^3$ in all our synthetic experiments. By the boundary setting as discussed in section \ref{sec:4_1}, the image of the resultant map is restricted to be the cube $\Omega$, even though the landmark moves towards the boundary.

\begin{figure}[t]
        \centering
        \begin{subfigure}[b]{0.33\textwidth}
                \includegraphics[width=\textwidth]{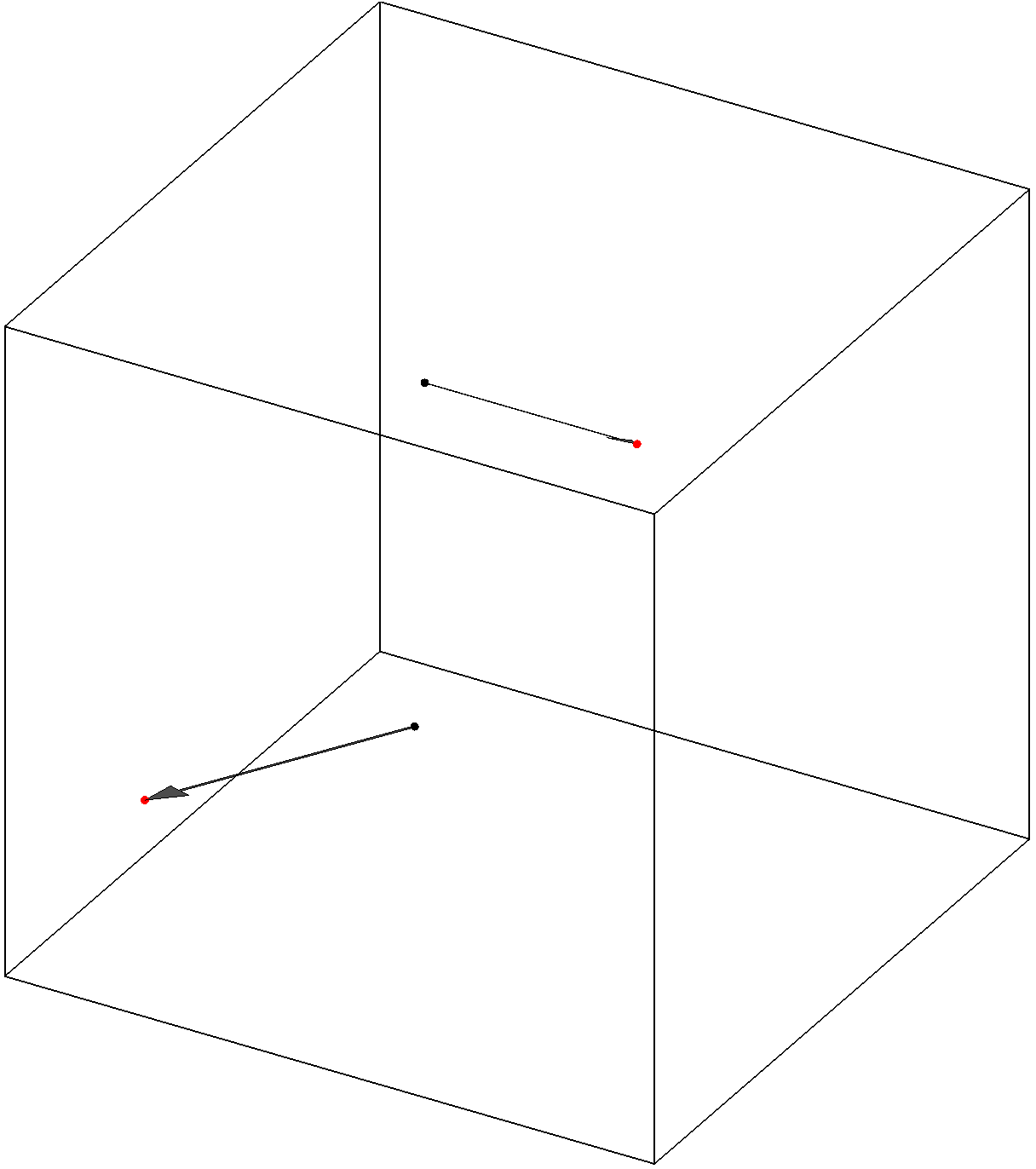}
                \caption{Landmark}
                \label{fig:two_pointA}
        \end{subfigure}
        \begin{subfigure}[b]{0.32\textwidth}
                \includegraphics[width=\textwidth]{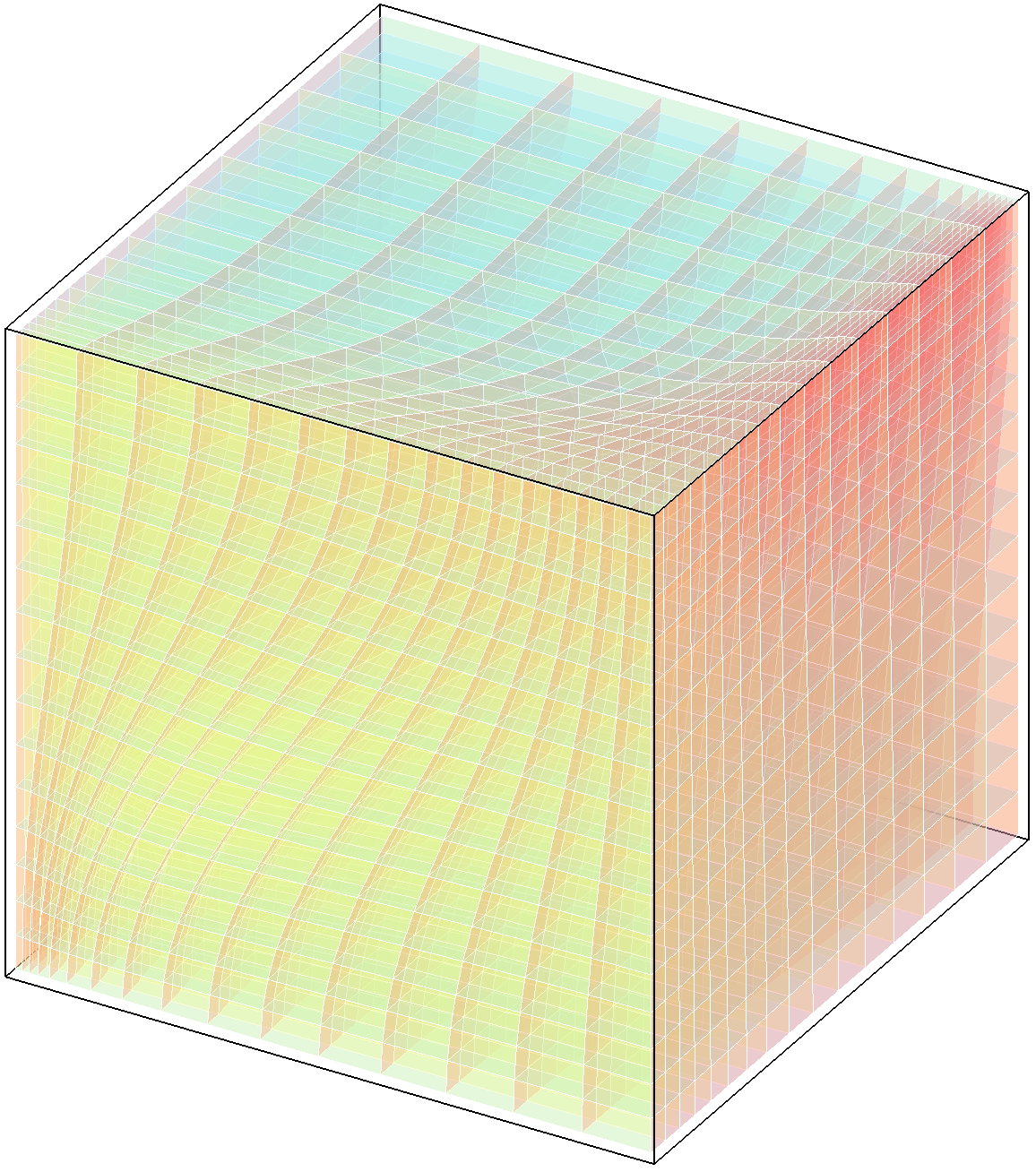}
                \caption{QC deformation}
                \label{fig:two_pointB}
        \end{subfigure}
        \begin{subfigure}[b]{0.32\textwidth}
                \includegraphics[width=\textwidth]{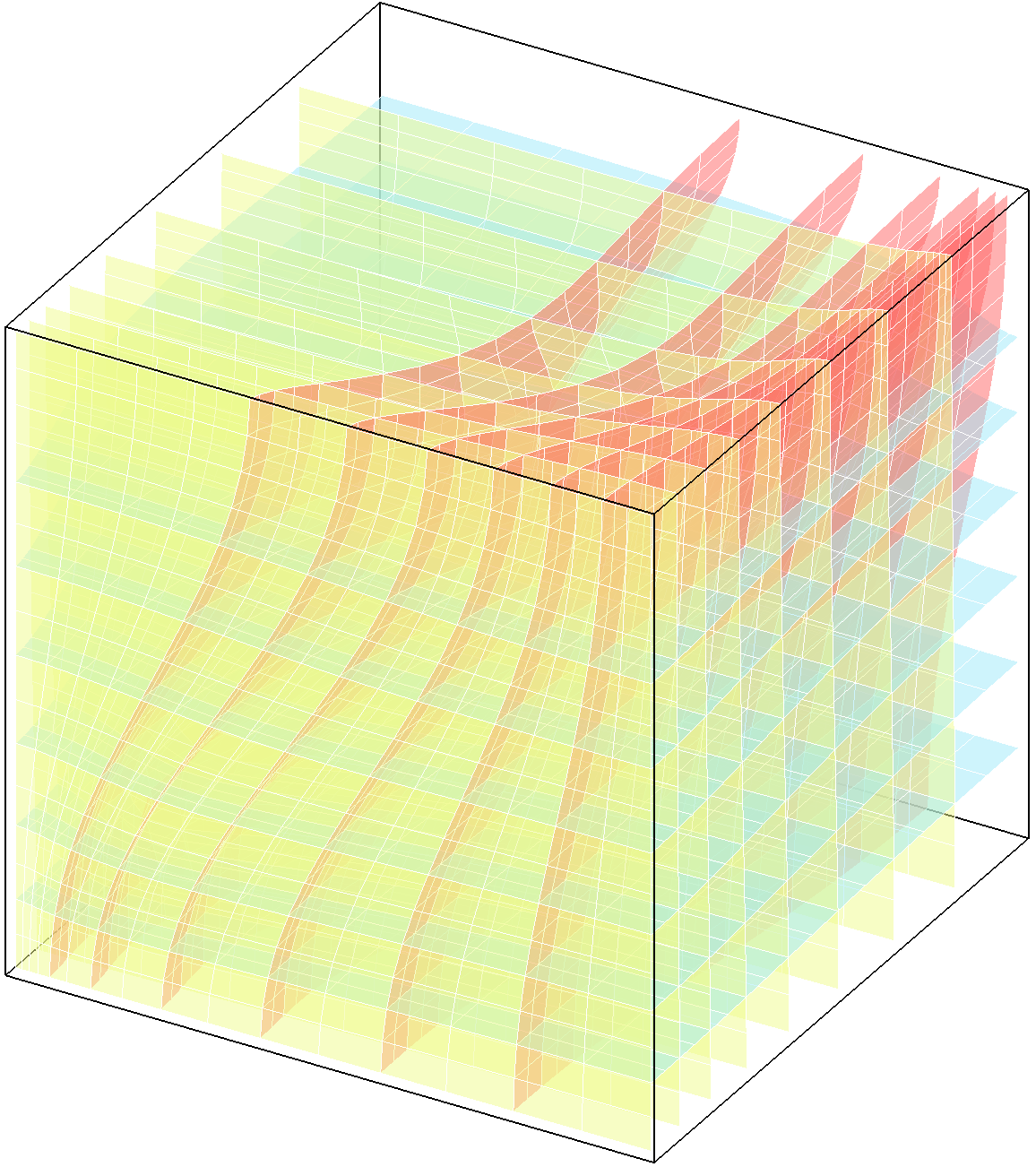}
                \caption{Sparse view}
                \label{fig:two_pointC}
        \end{subfigure}
\caption{Two-point landmark-matching experiment}
\end{figure}

Secondly, we test the algorithm to compute the landmark-matching transformation with two landmarks moving towards different directions. Deformations of both landmark points are large, as shown in Figure \ref{fig:two_pointA}. More specifically, two points $p_i$ are moved to $q_i = f(p_i)$ as follows:
\begin{equation}
\left[ \begin{array}{ccc}
p_1(x) & p_1(y) & p_1(z) \\
p_2(x) & p_2(y) & p_2(z) \end{array} \right] = \left[ \begin{array}{ccc}
0.6 & 0.7 & 0.7 \\
0.4 & 0.6 & 0.3 \end{array} \right] \rightarrow \left[ \begin{array}{ccc}
0.3 & 0.2 & 0.9 \\
0.2 & 0.9 & 0.2 \end{array} \right] = \left[ \begin{array}{ccc}
q_1(x) & q_1(y) & q_1(z) \\
q_2(x) & q_2(y) & q_2(z) \end{array} \right]
\end{equation}

Figure \ref{fig:two_pointB} shows the obtained transformation. Figure \ref{fig:two_pointC} shows the visualization of the obtained transformation with a sparser view.
\begin{figure}[t]
        \centering
        \begin{subfigure}[b]{0.33\textwidth}
                \includegraphics[width=\textwidth]{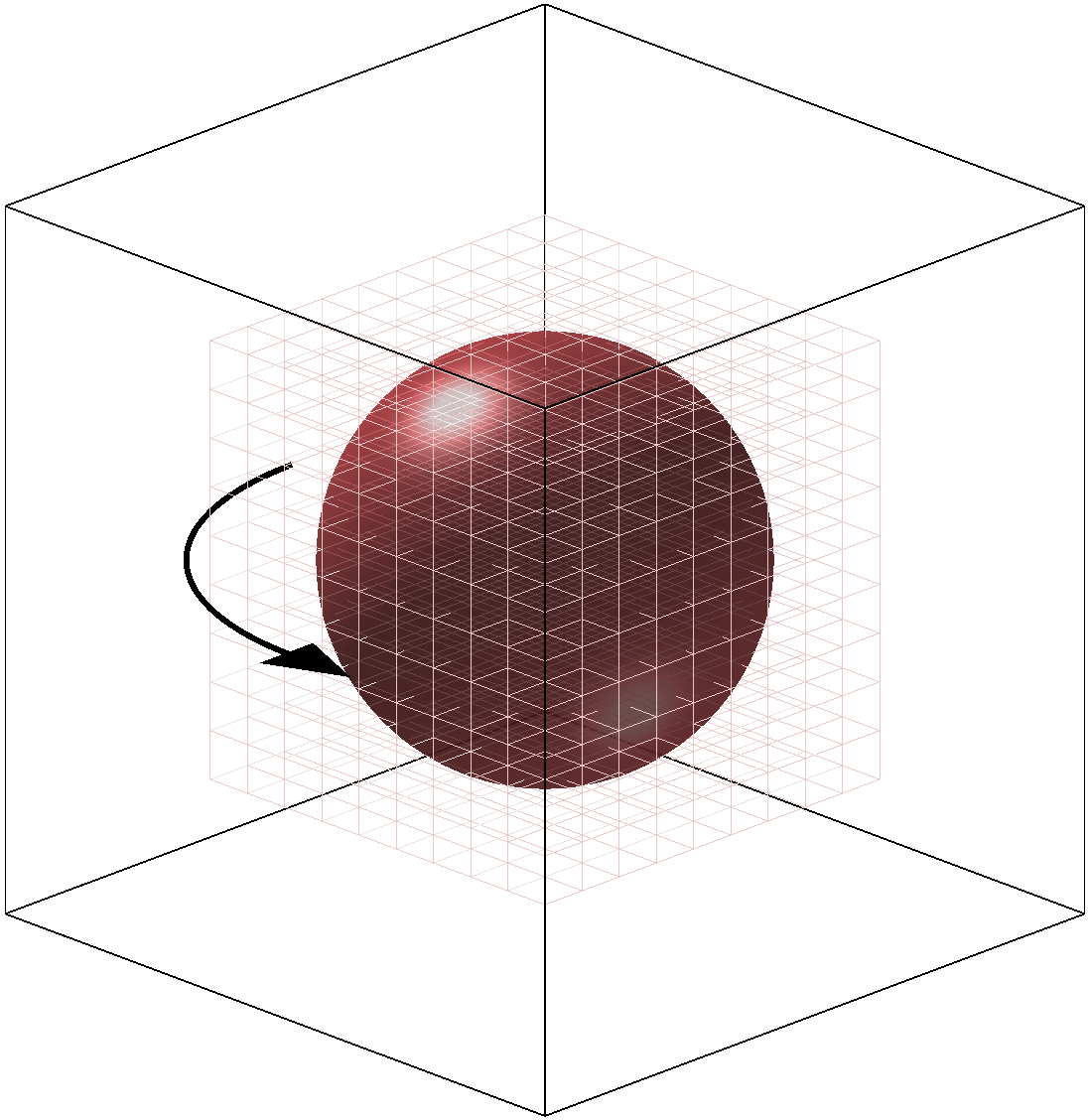}
                \caption{Landmark}
                \label{fig:rotateA}
        \end{subfigure}
        \begin{subfigure}[b]{0.32\textwidth}
                \includegraphics[width=\textwidth]{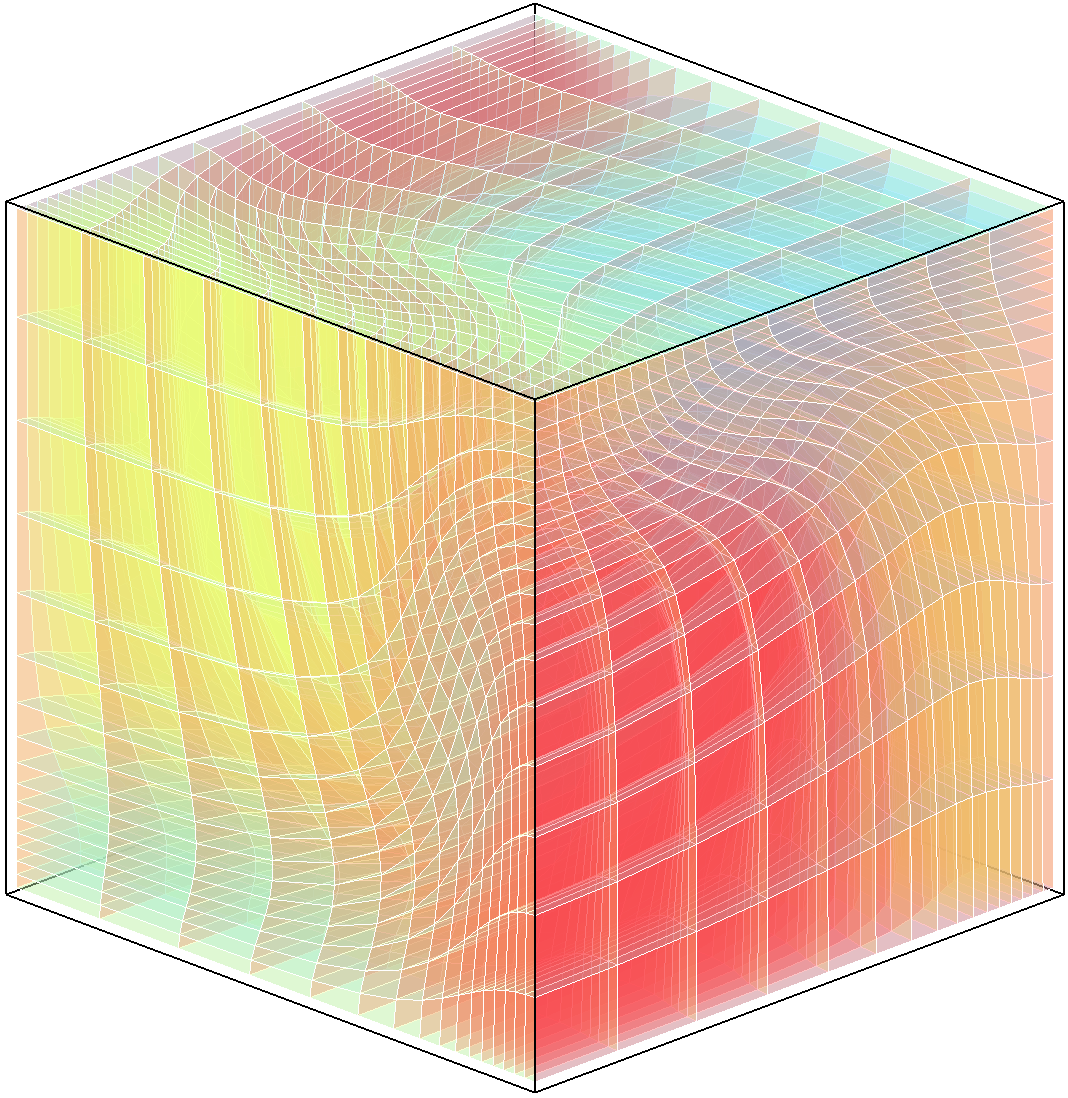}
                \caption{QC deformation}
                \label{fig:rotateB}
        \end{subfigure}
        \begin{subfigure}[b]{0.32\textwidth}
                \includegraphics[width=\textwidth]{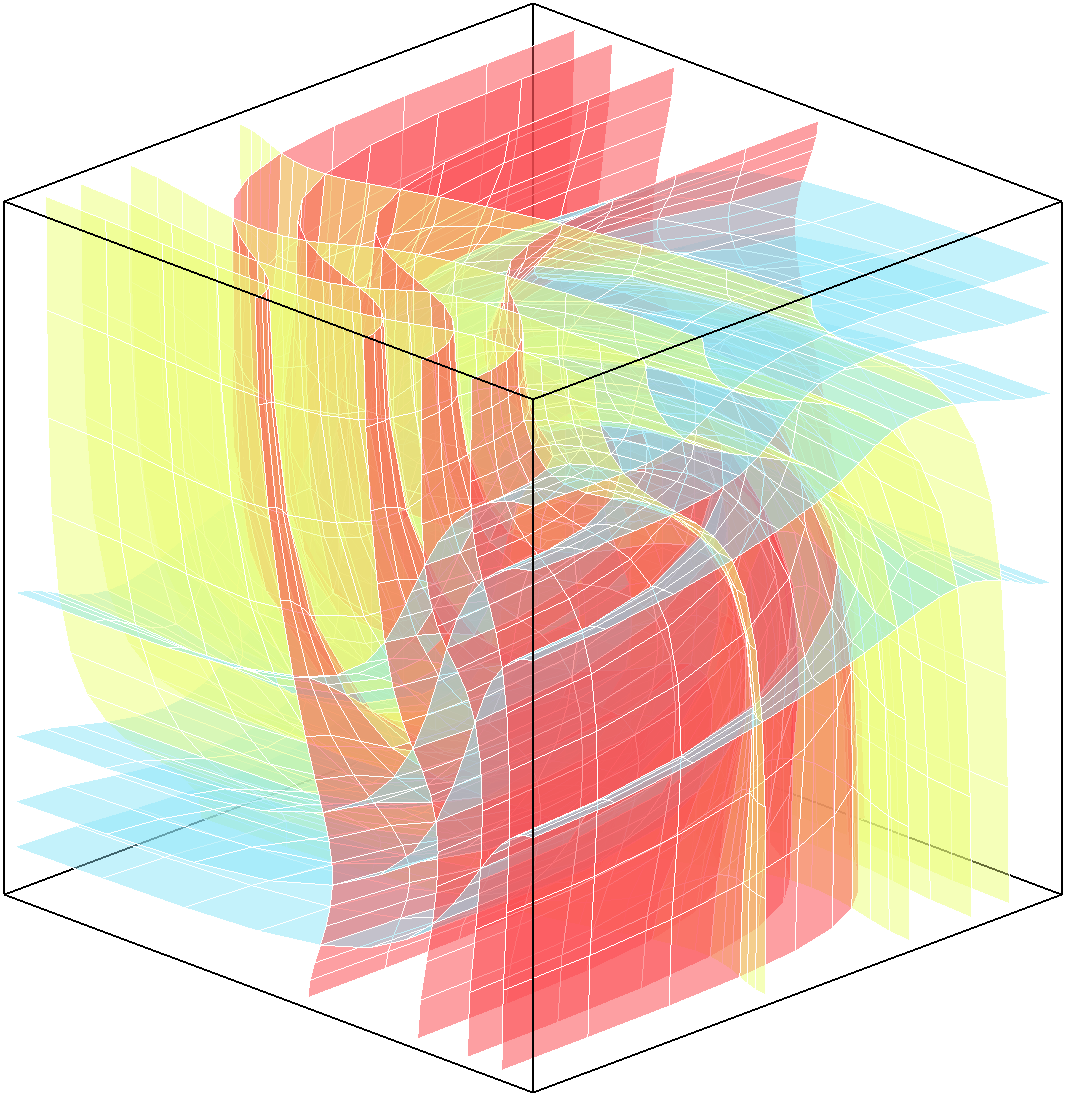}
                \caption{Sparse view}
                \label{fig:rotateC}
        \end{subfigure}
\caption{Landmark-matching experiment to register rotating ball.}
\label{fig:ball_rotate}
\end{figure}

We also test the algorithm to compute the landmark-matching transformation with an inner ball being chosen as landmarks. Points $p_i = (p_i(x), p_i(y), p_i(z))$ inside the inner ball are moved by the following transformation:
\begin{equation}
\left[ \begin{array}{c}
q_i(x) \\
q_i(y) \\
q_i(z) \end{array} \right] = \left[ \begin{array}{c}
f_1(p_i) \\
f_2(p_i) \\
f_3(p_i) \end{array} \right] = \left[ \begin{array}{ccc}
0 & -1 & 0 \\
1 & 0 & 0 \\
0 &  0 & 0 \end{array} \right]\left[ \begin{array}{c}
p_i(x) - 0.5 \\
p_i(y) - 0.5 \\
p_i(z) - 0.5 \end{array} \right]
\end{equation}
\noindent In other words, points inside the sphere are chosen as landmarks and they are rotated anti-clockwisely, as shown in Figure \ref{fig:rotateA}. The obtained  landmark-matching transformation, which is visualized as the deformation of the standard grid by the transformation, is shown in Figure \ref{fig:rotateB}. Figure \ref{fig:rotateC} visualizes the obtained transformation with a sparser view. Note that the obtained transformation is folding-free (Please refer to Table \ref{tab:Registration_result} which will be described later).
\begin{figure}[!h]
        \centering
        \begin{subfigure}[b]{0.33\textwidth}
                \includegraphics[width=\textwidth]{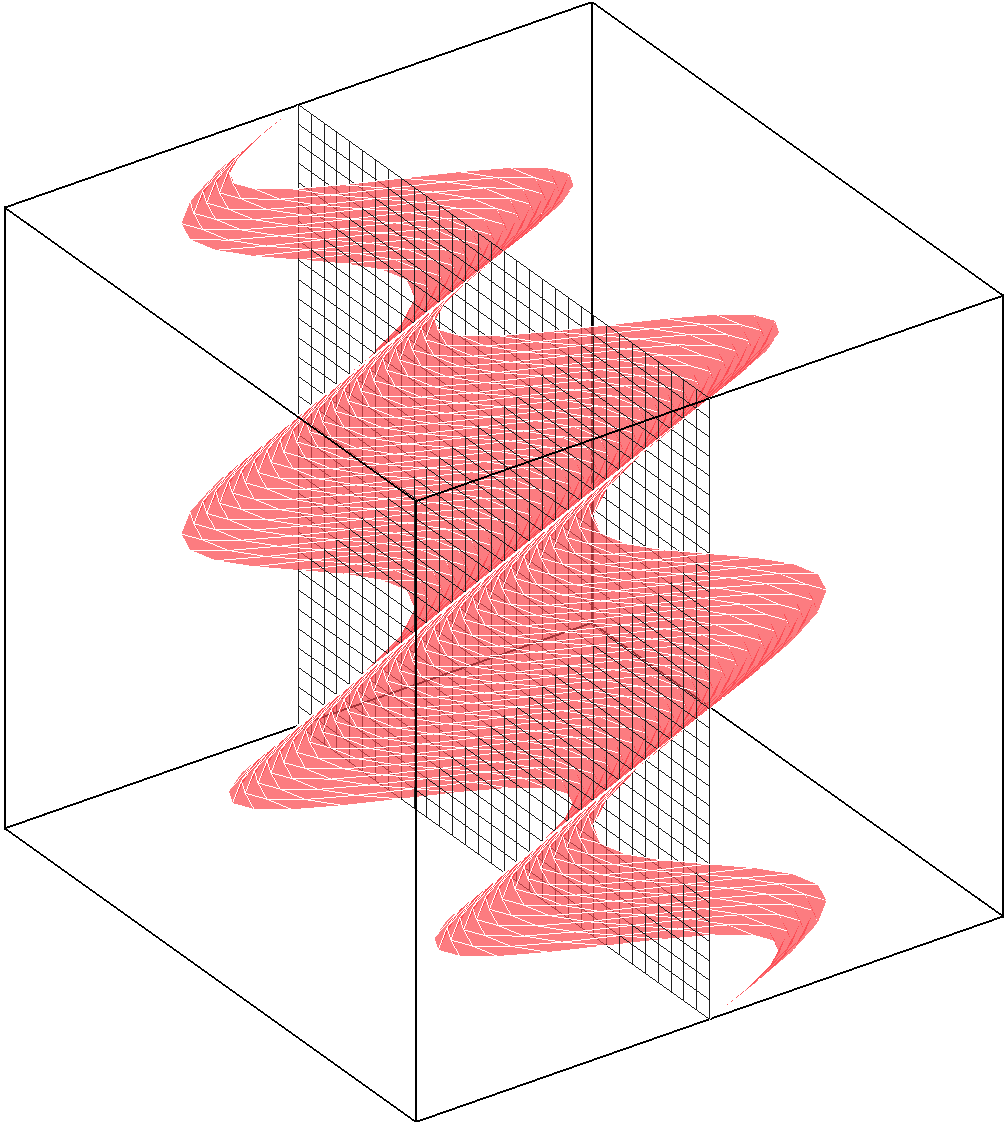}
	      \caption{Landmark}
                \label{fig:waveA}
        \end{subfigure}
        \begin{subfigure}[b]{0.32\textwidth}
                \includegraphics[width=\textwidth]{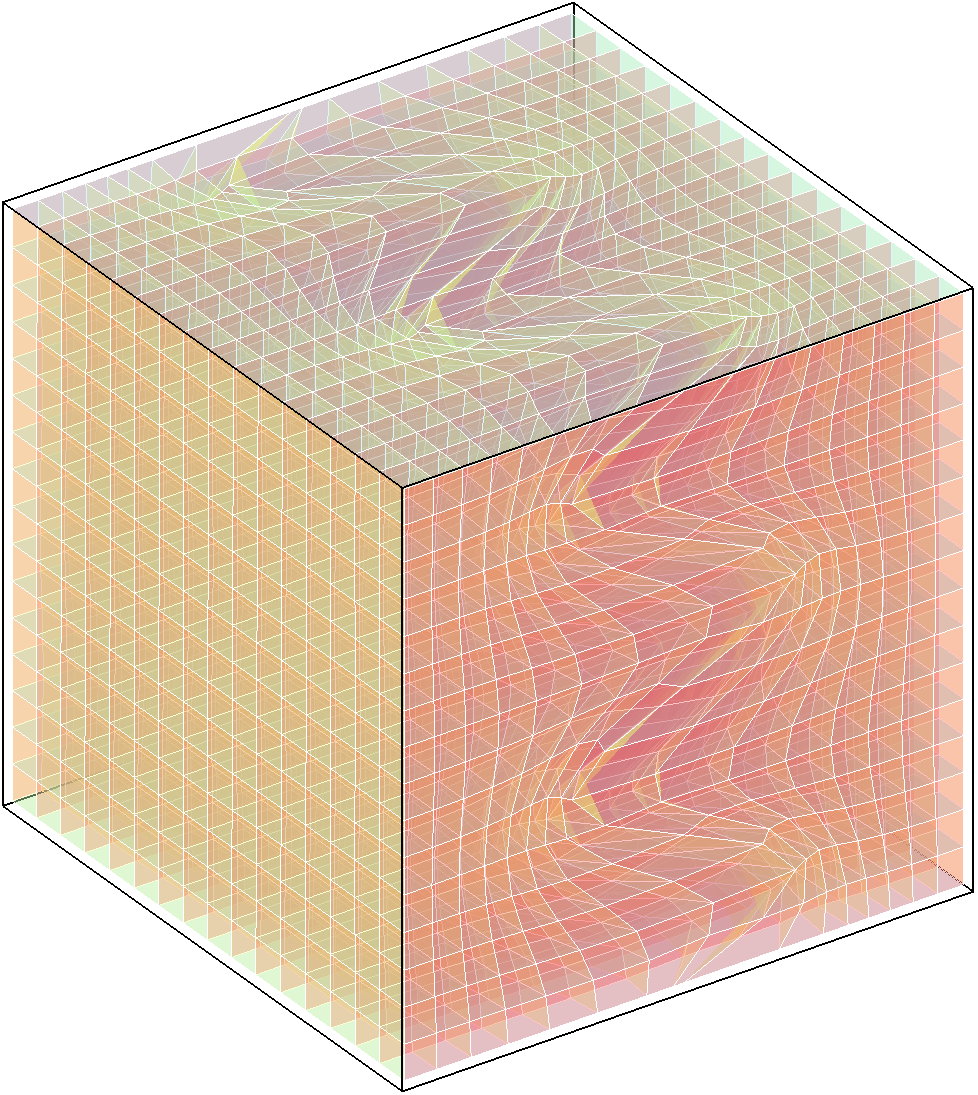}
	      \caption{QC deformation}
                \label{fig:waveB}
        \end{subfigure}
        \begin{subfigure}[b]{0.32\textwidth}
                \includegraphics[width=\textwidth]{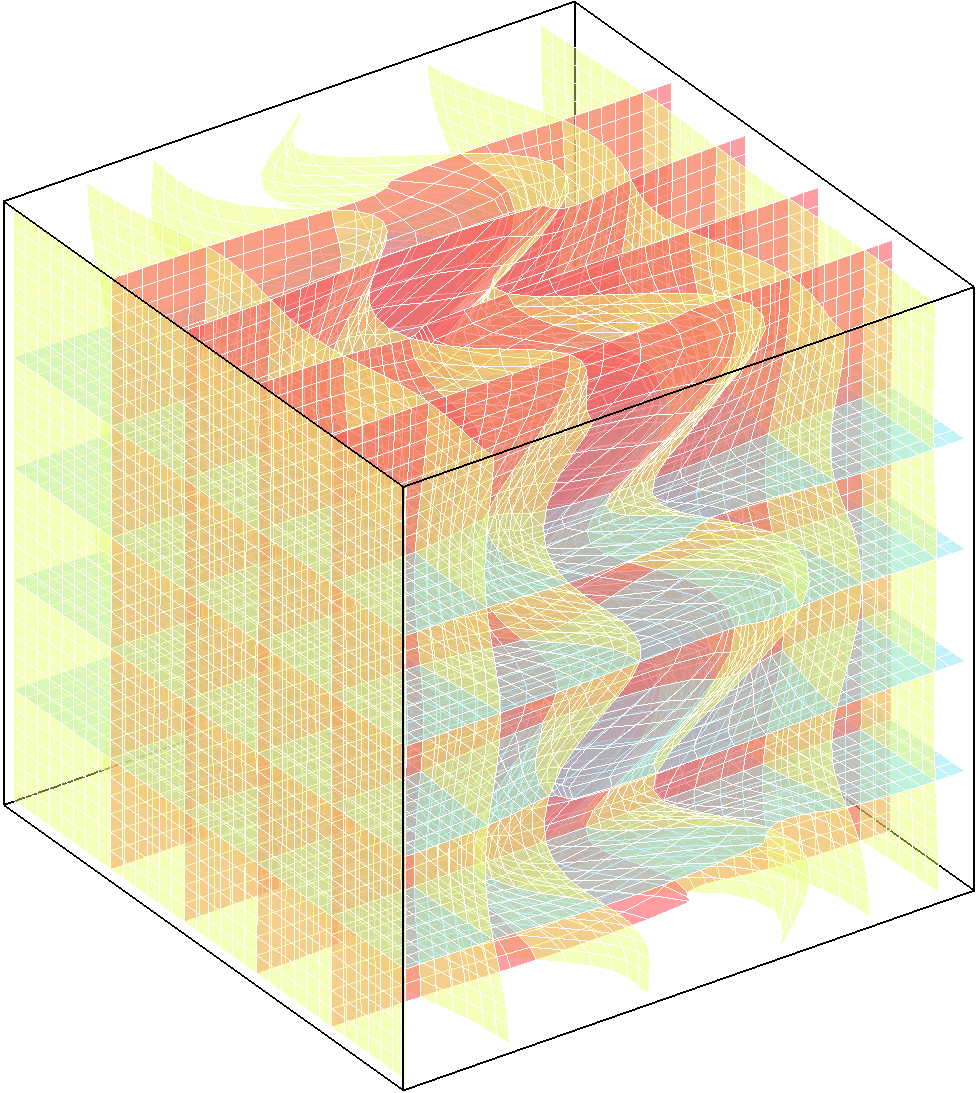}
	      \caption{Sparse view}
                \label{fig:waveC}
        \end{subfigure}
\caption{Landmark-matching experiment to register wave deformation}
\end{figure}

Next, we test the algorithm on an example of which all the points on a plane 
\[
p_i = [ 0.5, p_i(y), p_i(z) ] \ \forall (p_i(y),p_i(z)) \in [0,1]\times [0,1]
\]
are chosen as landmarks (grey plane in Figure \ref{fig:waveA}). The landmarks are deformed to a wave-shape surface (red surface in Figure \ref{fig:waveB}) by the following transformation:
\begin{equation}
\left[ \begin{array}{c}
q_i(x) \\
q_i(y) \\
q_i(z) \end{array} \right] = \left[ \begin{array}{c}
f_1(p_i) \\
f_2(p_i) \\
f_3(p_i) \end{array} \right] = \left[ \begin{array}{c}
0.5 + \frac{1}{5}\sin\left(4\pi(p_i(y) + p_i(z))\right) \\
p_i(y) \\
p_i(z) \end{array} \right].
\end{equation}
Using our proposed algorithm, we obtain a transformation that satisfies the landmark constraints. Figure \ref{fig:waveB} shows the obtained transformation, which is bijective (Refer to Table \ref{tab:Registration_result}). Figure \ref{fig:waveC} shows the transformation with a sparser view.

\begin{figure}[t]
        \centering
        \begin{subfigure}[b]{0.33\textwidth}
                \includegraphics[width=\textwidth]{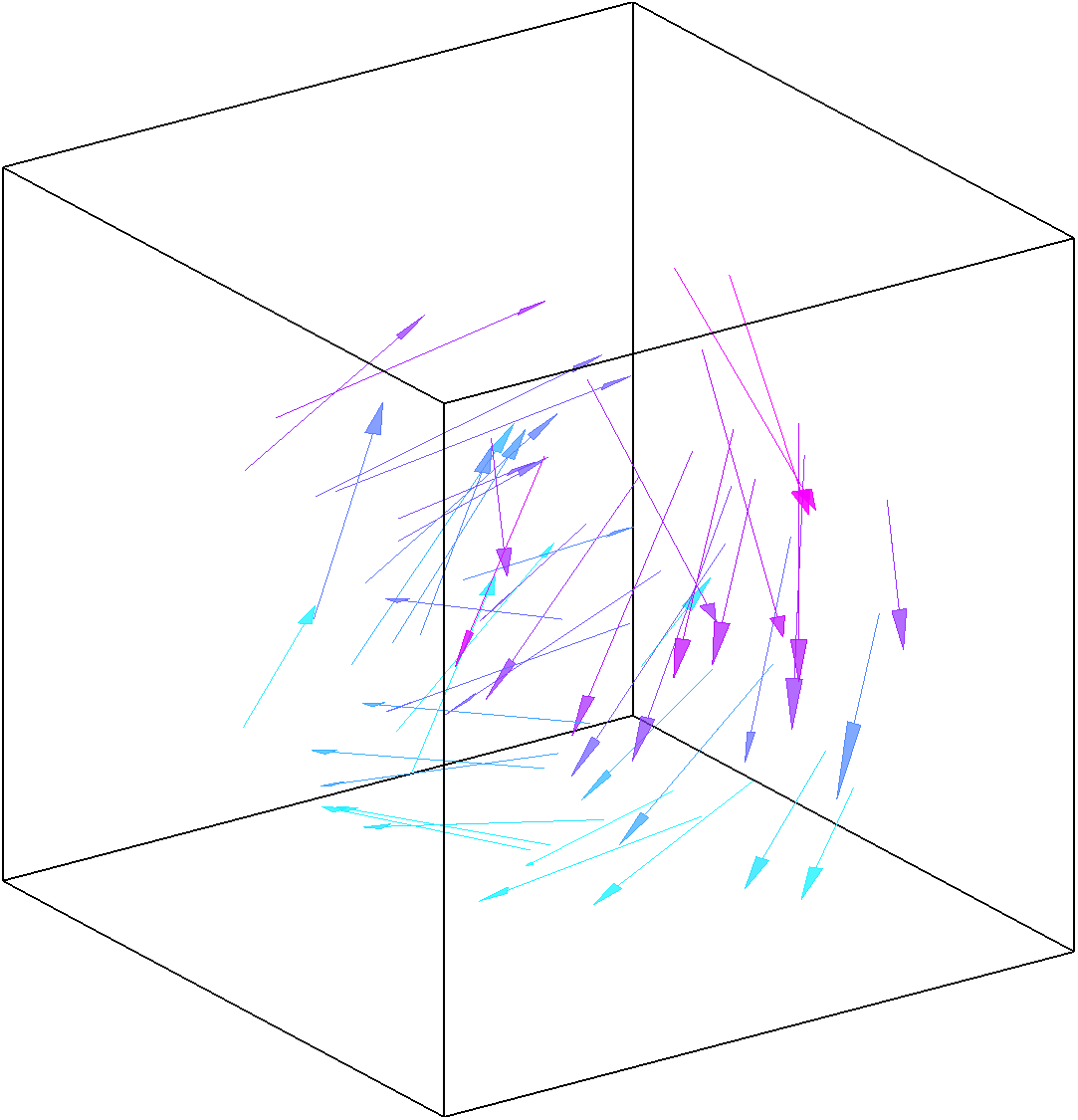}
	      \caption{Landmark}
                \label{fig:twistA}
        \end{subfigure}
        \begin{subfigure}[b]{0.32\textwidth}
                \includegraphics[width=\textwidth]{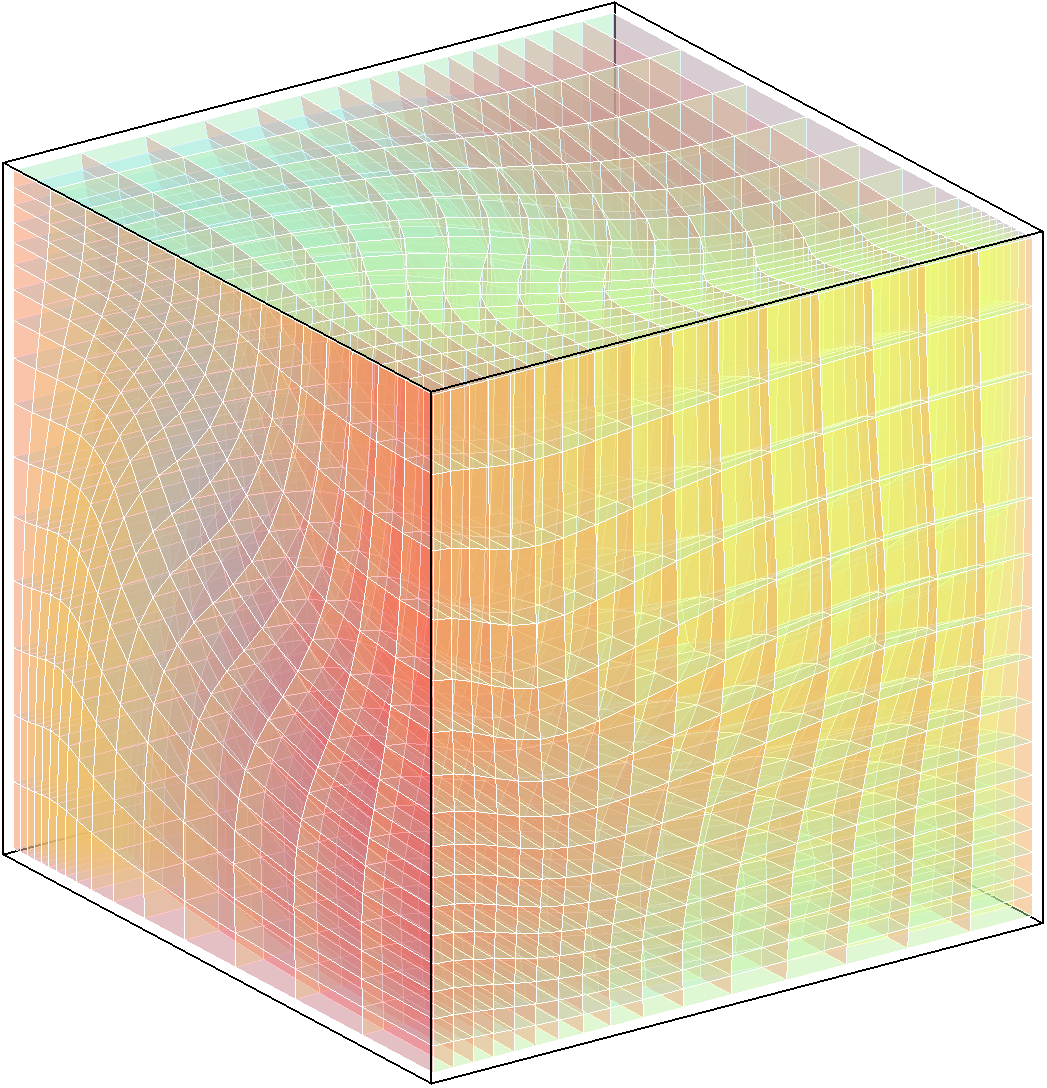}
	      \caption{QC deformation}
                \label{fig:twistB}
        \end{subfigure}
        \begin{subfigure}[b]{0.32\textwidth}
                \includegraphics[width=\textwidth]{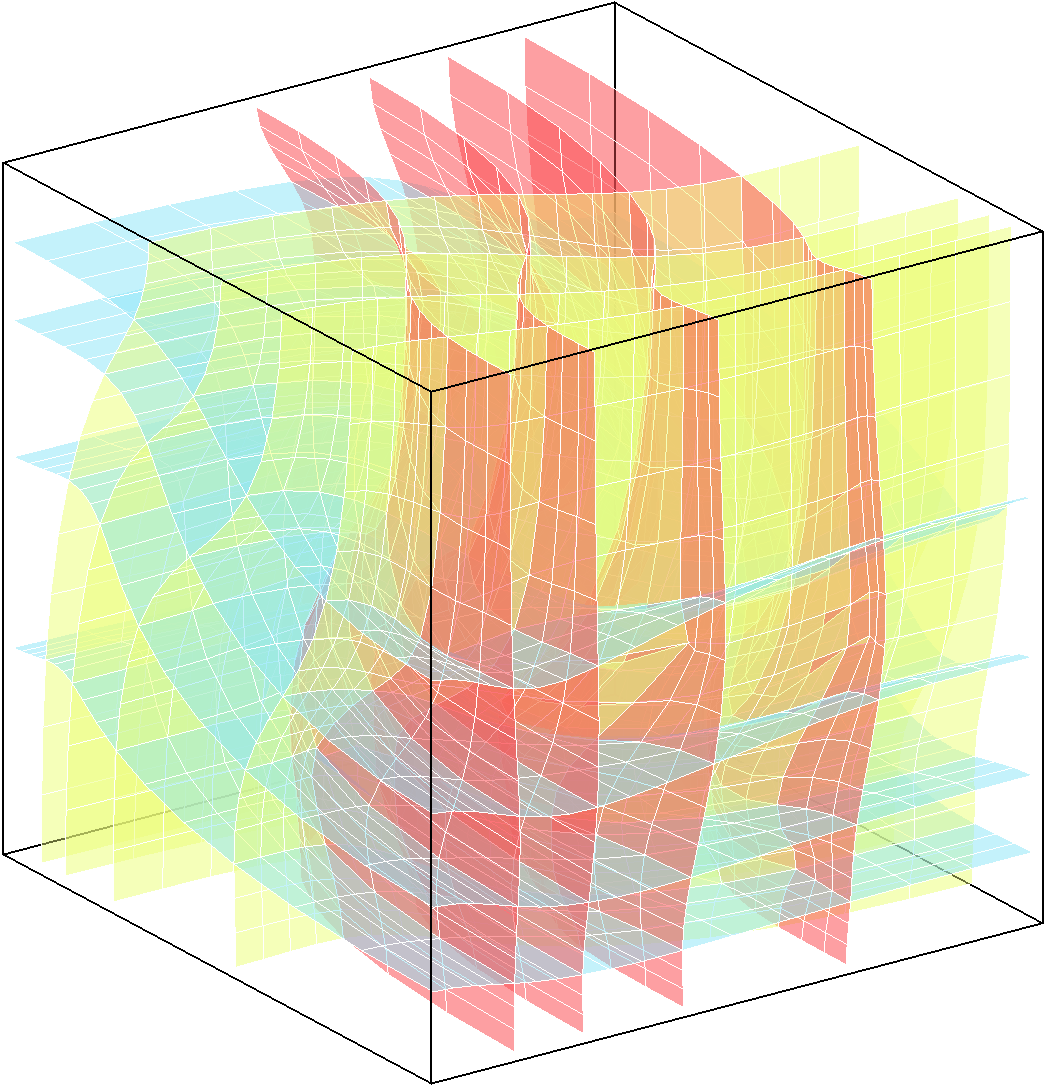}
	      \caption{Sparse view}
                \label{fig:twistC}
        \end{subfigure}
\caption{Landmark-matching experiment of random points with twisting deformation}
\end{figure}
Finally, we test the algorithm to compute the landmark-matching transformation with random points being chosen as landmarks. These random landmark points are twisted by the following transformation (See Figure \ref{fig:twistA}):
\begin{equation}
\begin{array}{lll}
q_i(x) &=& f_1(p_i) = p_i(x) - p_i(x) \cdot A\left(p_i(x),p_i(y)\right), \\
q_i(y) &=& f_2(p_i) = \rho - p_i(z) \cdot A\left( \rho, p_i(z) \right), \\
q_i(z) &=& f_3(p_i) = p_i(z) + q_i(y) \cdot A\left( \rho, p_i(z) \right).
\end{array}
\end{equation}
\noindent where
\begin{equation}
\begin{array}{c}
\rho =  p_i(y) + q_i(x) \cdot A\left(p_i(x),p_i(y)\right) \\
A(x,y) = \frac{1}{100}\left( \frac{\left(\cos(\pi x) + 1\right)\left(\cos(\pi y) + 1\right)}{4} + \cos\left(\frac{\pi x}{2}\right)\cos\left(\frac{\pi y}{2}\right)\right)
\end{array}
\end{equation}
The twisting deformation is large and complicated. Using our algorithm, we are able to obtain a diffeomorphic landmark-matching transformation. Figure \ref{fig:twistB} shows the obtained transformation. Figure \ref{fig:twistC} shows the registration with a sparser view.

\begin{figure}[t]
        \centering
        \begin{subfigure}[b]{0.32\textwidth}
                \includegraphics[width=\textwidth]{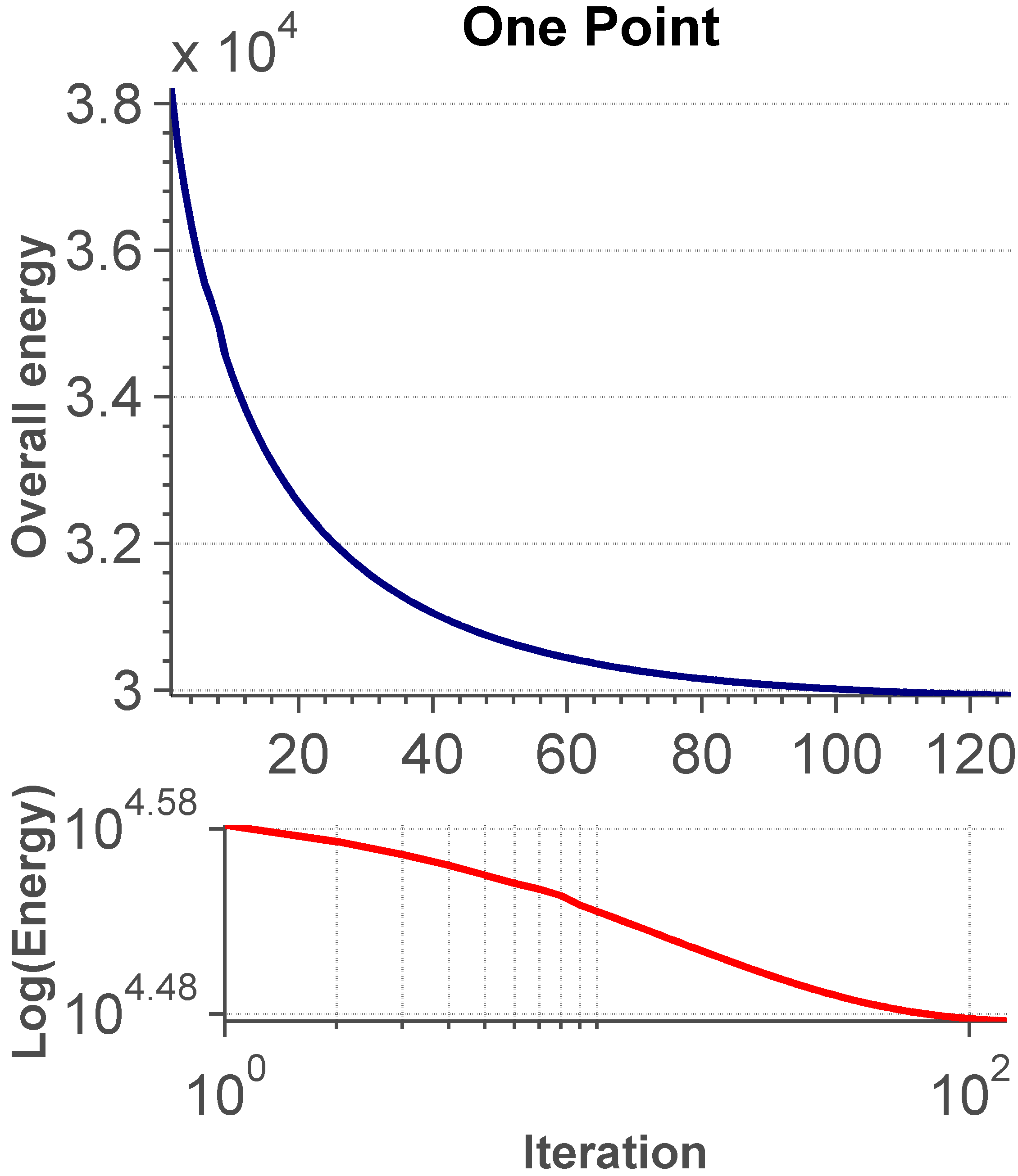}
                \caption{One-point}
        \end{subfigure}
        \begin{subfigure}[b]{0.32\textwidth}
                \includegraphics[width=\textwidth]{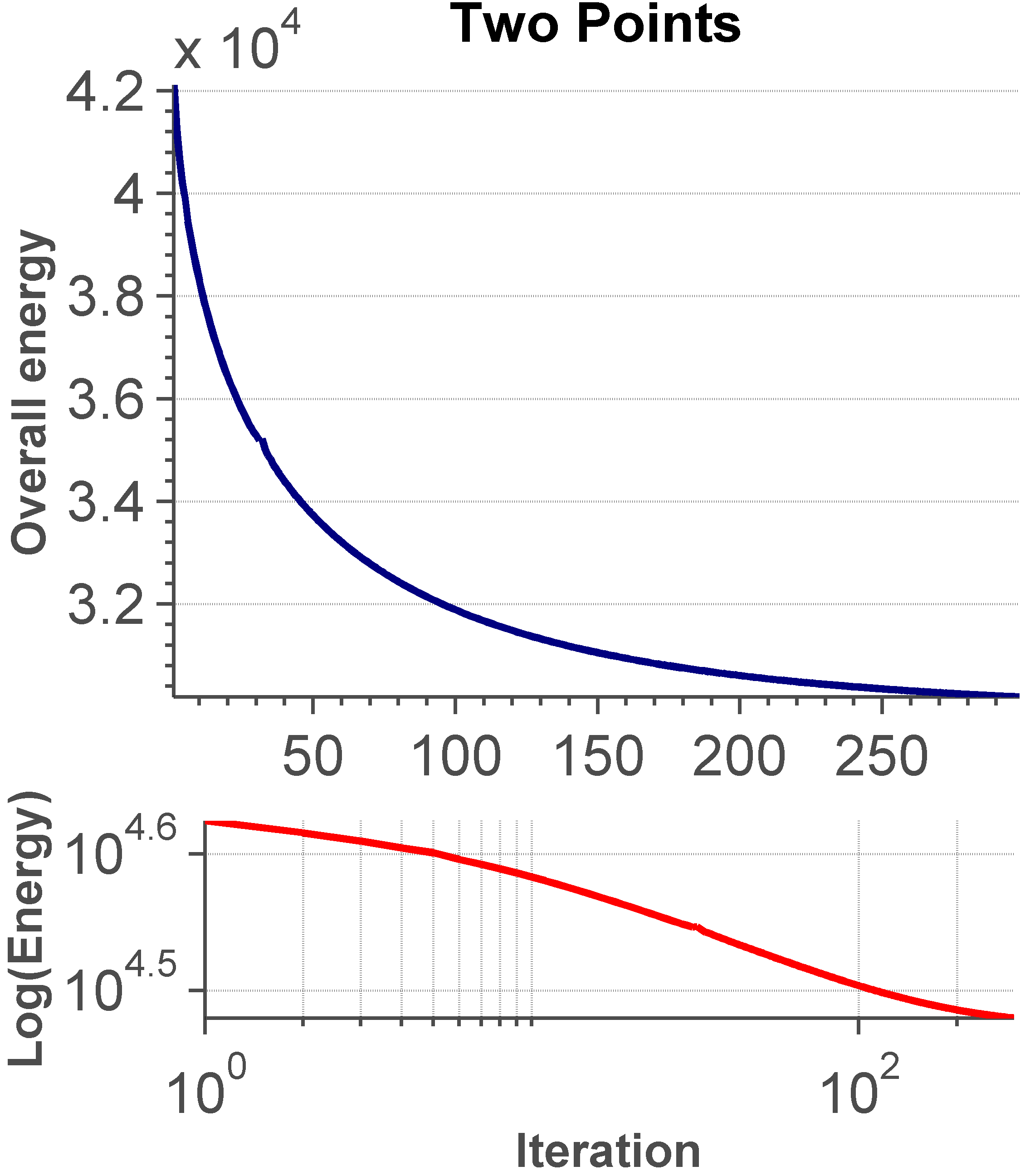}
                \caption{Two-point}
        \end{subfigure}
        \begin{subfigure}[b]{0.32\textwidth}
                \includegraphics[width=\textwidth]{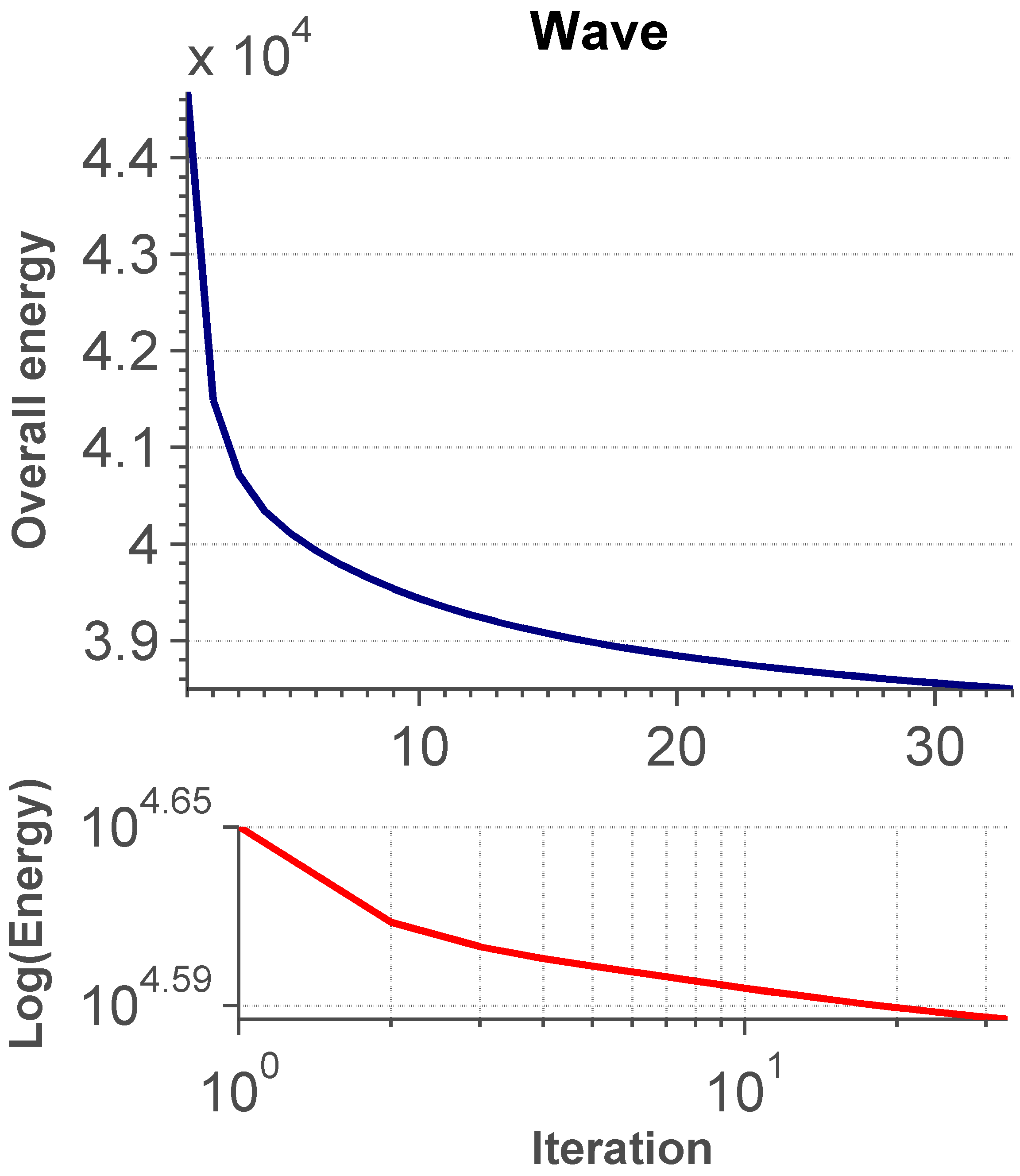}
                \caption{Wave}
        \end{subfigure}
        \begin{subfigure}[b]{0.32\textwidth}
                \includegraphics[width=\textwidth]{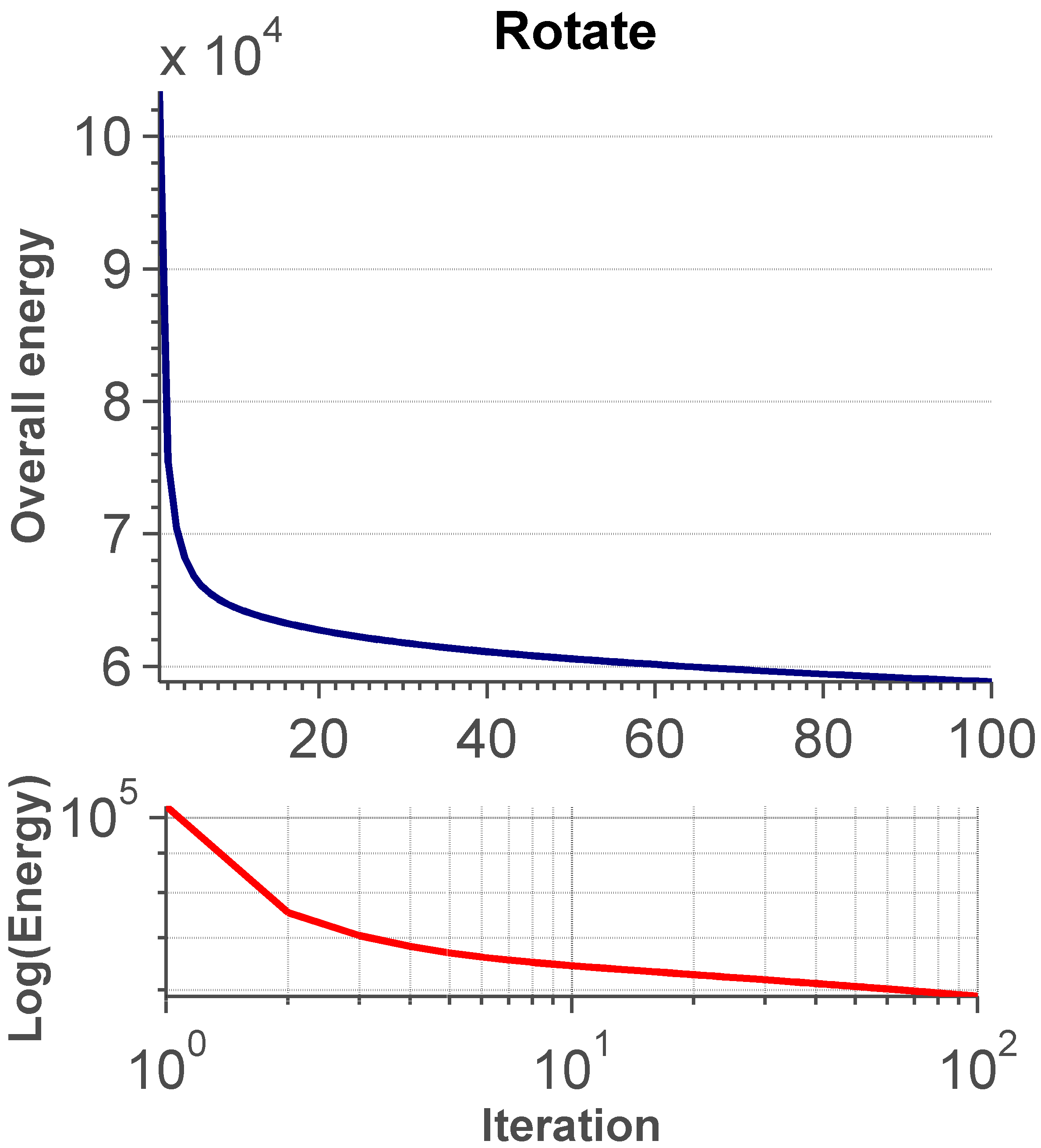}
                \caption{Rotate}
        \end{subfigure}
        \begin{subfigure}[b]{0.32\textwidth}
                \includegraphics[width=\textwidth]{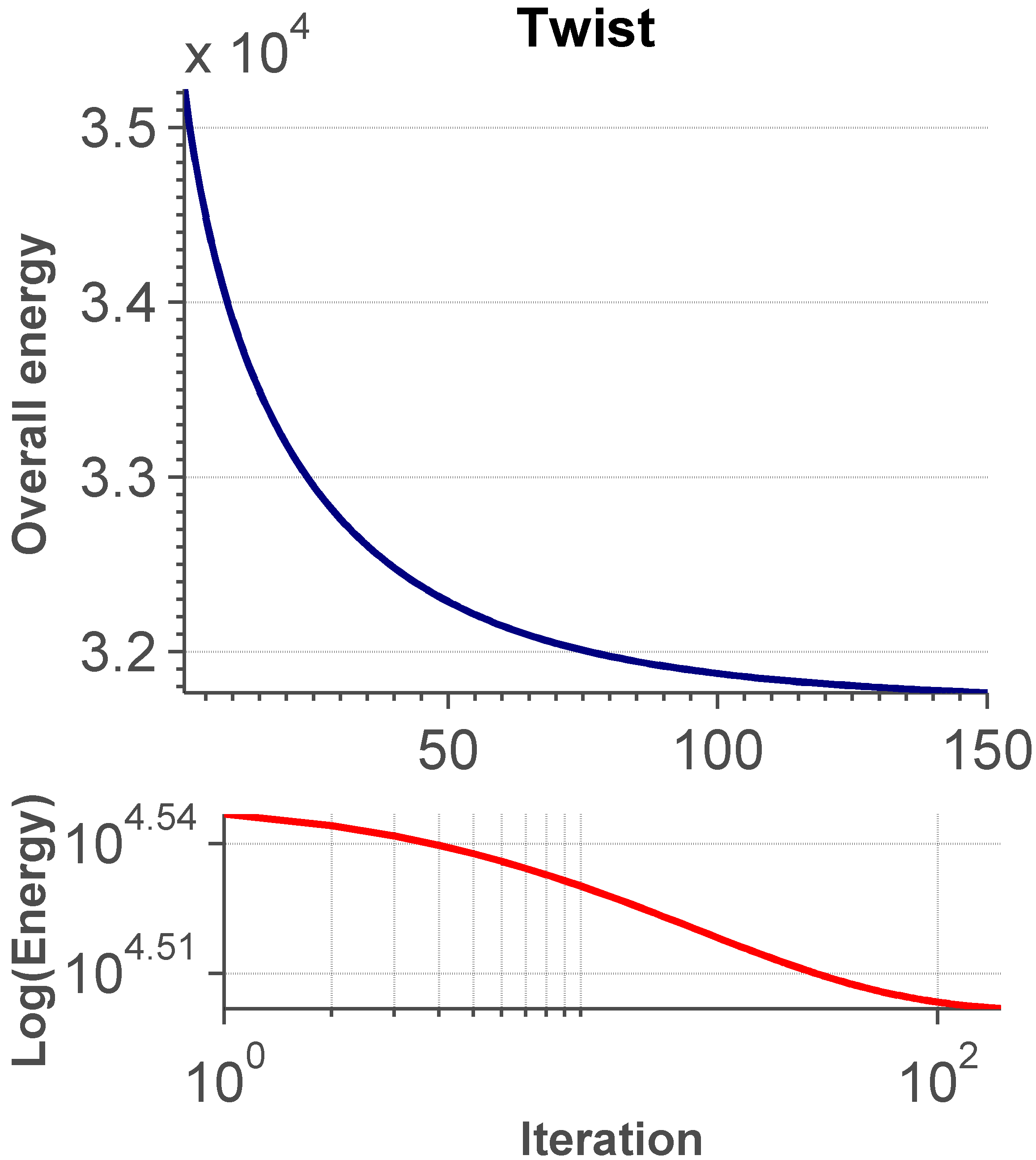}
                \caption{Twist}
        \end{subfigure}
        \caption{The overall energy versus iterations.\label{fig:energy}}
\end{figure}

The upper row of figure \ref{fig:energy}(a)--(e) shows the overall energy (See (\ref{eq:our_energy}) versus iterations for the ``one point landmark", ``two-point landmark", ``wave-shape deformation", ``rotate sphere" and ``twist point sets" examples respectively. Note that the overall energy of each mapping is iteratively reduced with a trend of converging to an optimal map with respect to our proposed model \ref{eq:our_energy}. The second row shows the corresponding log-log plot of the overall energy versus iterations. The negative slope appear in all five examples indicates that our proposed algorithm successfully minimizes the generalized conformality distortion $K(f)$ while matching the prescribed landmark correspondences.

The above examples demonstrate that our proposed algorithm is effective for computing landmark-matching folding-free transformation with larger deformations. It works well even with large number of landmarks or large deformations.

\subsection*{Lung CT landmark-based image registration}
\begin{figure}[t]
        \centering
        \begin{subfigure}[b]{0.32\textwidth}
                \includegraphics[width=\textwidth]{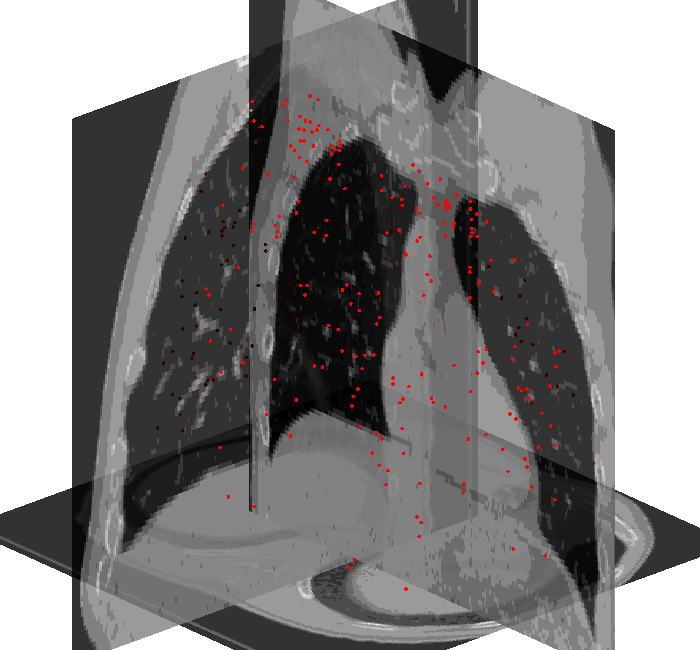}
                \caption{Lung at time = 00}
                \label{fig:Lung1A}
        \end{subfigure}
        \begin{subfigure}[b]{0.32\textwidth}
                \includegraphics[width=\textwidth]{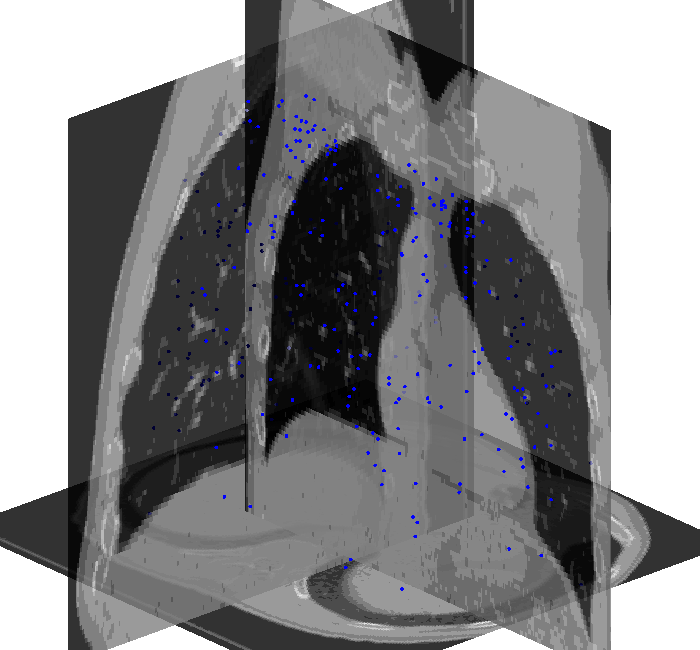}
                \caption{Lung at time = 50}
                \label{fig:Lung1B}
        \end{subfigure}
        \begin{subfigure}[b]{0.32\textwidth}
                \includegraphics[width=\textwidth]{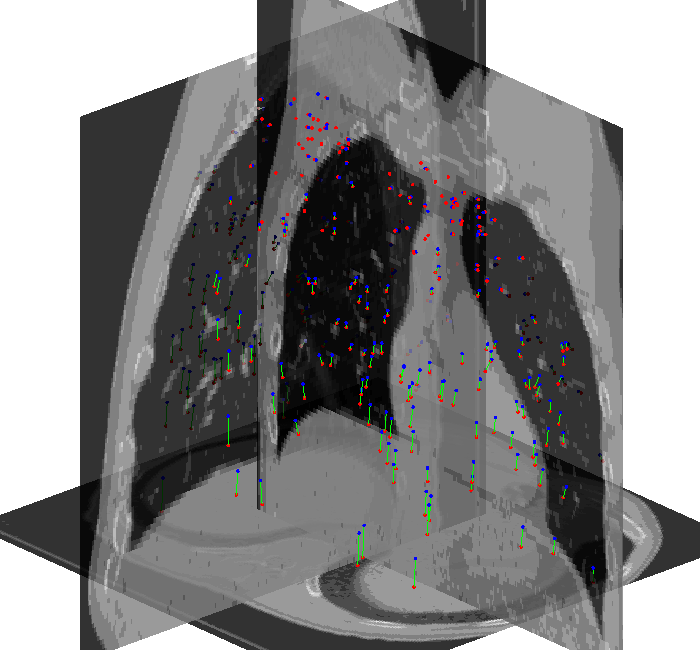}
                \caption{Registration result}
                \label{fig:Lung1_register}
        \end{subfigure}
        \caption{Lung CT image registration (CT 1)}\label{fig:Lung1_3D}
\end{figure}

We have also applied our algorithm to compute landmark-matching transformation of real four dimensional lung CT data with prescribed landmark correspondences at different times. Five sets of lung CT images are registered using our proposed algorithm. We choose the maximum inhalation phase image (at time $t = 00$) and the maximum exhalation phase image (at time $t = 50$) as the moving image and the reference image respectively. This provides the maximum displacement of the landmarks located within the lung CT images. To demonstrate the independence of our algorithm to the number of landmark points, 300 prescribed feature correspondences are enforced. Figure \ref{fig:Lung1A} and \ref{fig:Lung1B} show the lung CT images at time $t=00$ and $t=50$. The image dimension of this dataset is $256\times 256\times 112$. Since the multi-grid method is applied to obtain a preconditioner to solve the f-subproblem, a linear interpolation on the image is firstly done to get the position of the landmarks corresponding to the dimension $256\times 256 \times 128$, in which every dimensions has grid spacing equals to the power of 2. The 300 prescribed landmark correspondences between the two images are shown as the red and blue dots in the figures. Using the proposed algorithm, the landmark-based image registration of the lung CT images can be computed, which is shown in Figure \ref{fig:Lung1_register}.
\begin{figure}[t]
        \centering
        \begin{subfigure}[b]{0.32\textwidth}
                \includegraphics[width=\textwidth]{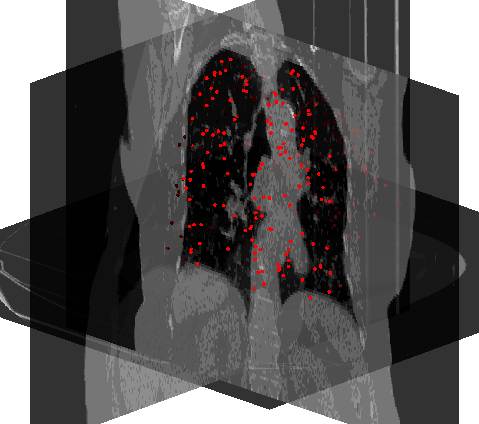}
                \caption{Lung at time = 00}
                \label{fig:Lung2A}
        \end{subfigure}
        \begin{subfigure}[b]{0.32\textwidth}
                \includegraphics[width=\textwidth]{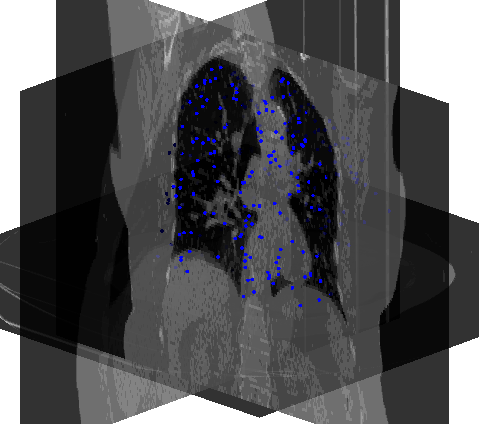}
                \caption{Lung at time = 50}
                \label{fig:Lung2B}
        \end{subfigure}
        \begin{subfigure}[b]{0.32\textwidth}
                \includegraphics[width=\textwidth]{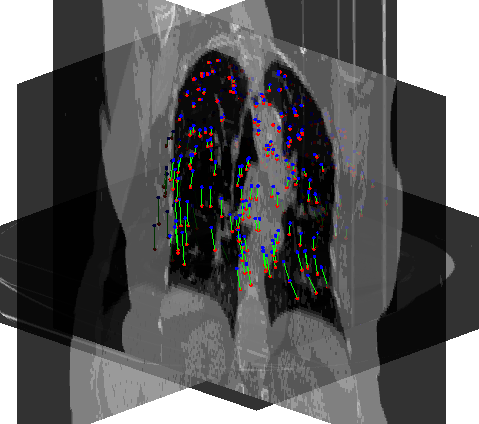}
                \caption{Registration result}
                \label{fig:Lung2_register}
        \end{subfigure}
        \caption{Lung CT image registration (CT 5)}\label{fig:Lung2_3D}
\end{figure}

Figure \ref{fig:Lung2A} and \ref{fig:Lung2B} show another set of lung CT images at time $t=00$ and $t=50$. The image dimension of this dataset is $512\times 512\times 128$. The 300 prescribed landmark correspondences are shown as the red and blue dots in the figures. The obtained landmark-based image registration of the lung CT images is shown in Figure \ref{fig:Lung2_register}.

\begin{figure}
        \centering
        \begin{subfigure}[b]{0.32\textwidth}
                \includegraphics[width=\textwidth]{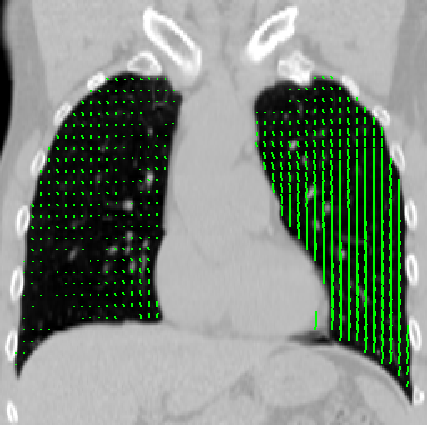}
                \caption{X-slide 50}
        \end{subfigure}
        \begin{subfigure}[b]{0.32\textwidth}
                \includegraphics[width=\textwidth]{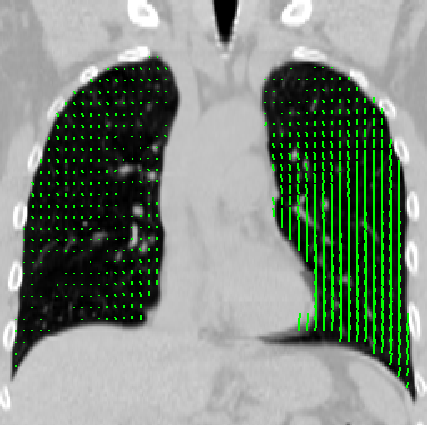}
                \caption{X-slide 55}
        \end{subfigure}
        \begin{subfigure}[b]{0.32\textwidth}
                \includegraphics[width=\textwidth]{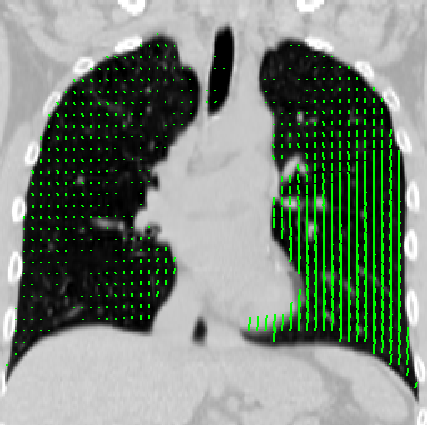}
                \caption{X-slide 60}
        \end{subfigure}
        \begin{subfigure}[b]{0.32\textwidth}
                \includegraphics[width=\textwidth]{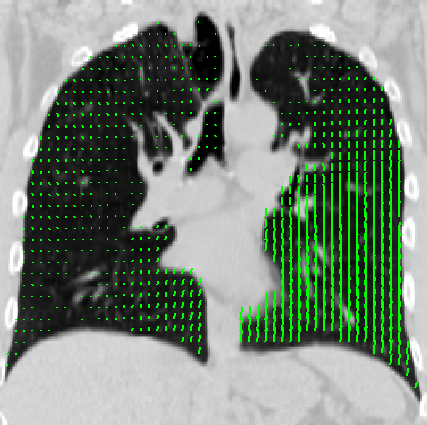}
                \caption{X-slide 65}
        \end{subfigure}
        \begin{subfigure}[b]{0.32\textwidth}
                \includegraphics[width=\textwidth]{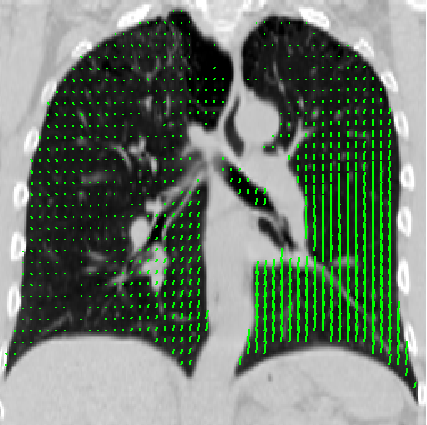}
                \caption{X-slide 70}
        \end{subfigure}
        \begin{subfigure}[b]{0.32\textwidth}
                \includegraphics[width=\textwidth]{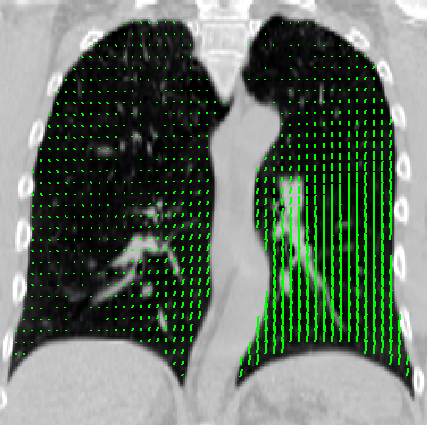}
                \caption{X-slide 75}
        \end{subfigure}
        \caption{Vector field of the registration result (CT1)}
\label{fig:X_slide_lung1}
\end{figure}

\begin{figure}
        \centering
        \begin{subfigure}[b]{0.32\textwidth}
                \includegraphics[width=\textwidth]{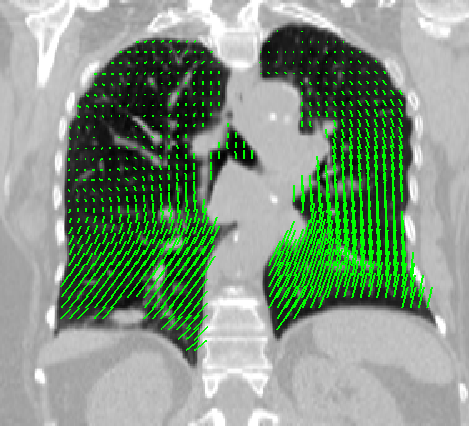}
                \caption{X-slide 50}
        \end{subfigure}
        \begin{subfigure}[b]{0.32\textwidth}
                \includegraphics[width=\textwidth]{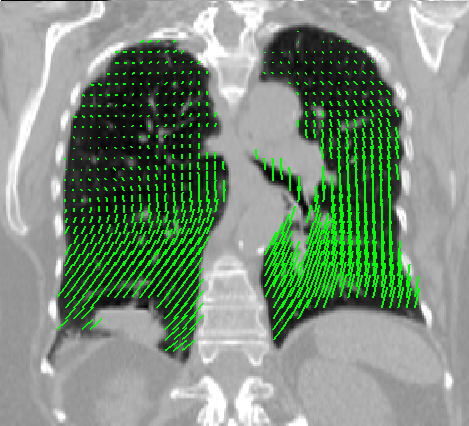}
                \caption{X-slide 55}
        \end{subfigure}
        \begin{subfigure}[b]{0.32\textwidth}
                \includegraphics[width=\textwidth]{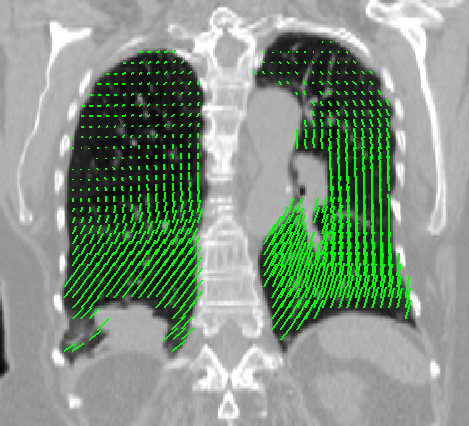}
                \caption{X-slide 60}
        \end{subfigure}
        \begin{subfigure}[b]{0.32\textwidth}
                \includegraphics[width=\textwidth]{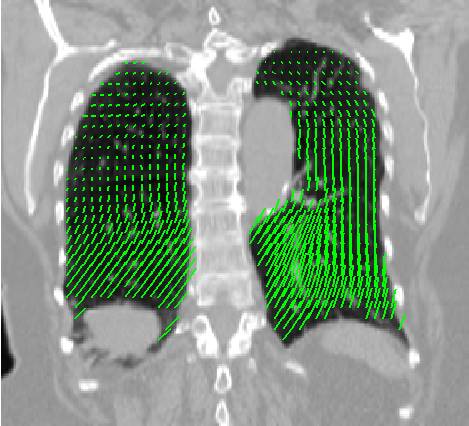}
                \caption{X-slide 65}
        \end{subfigure}
        \begin{subfigure}[b]{0.32\textwidth}
                \includegraphics[width=\textwidth]{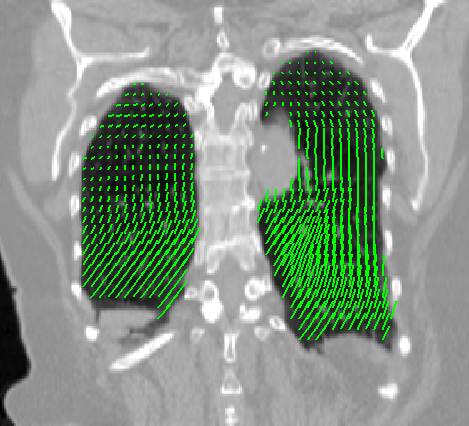}
                \caption{X-slide 70}
        \end{subfigure}
        \begin{subfigure}[b]{0.32\textwidth}
                \includegraphics[width=\textwidth]{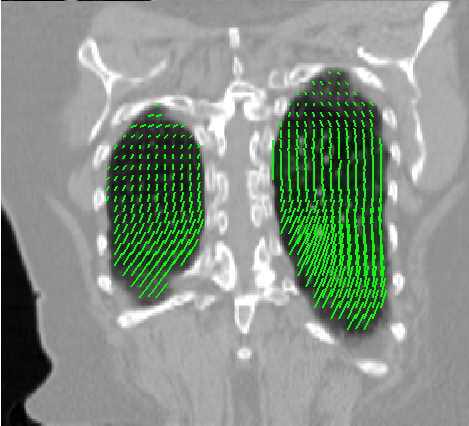}
                \caption{X-slide 75}
        \end{subfigure}
        \caption{Vector field of the registration result (CT5)}
\label{fig:X_slide_lung2}
\end{figure}

Figure \ref{fig:X_slide_lung1}(a)--(f) and \ref{fig:X_slide_lung2}(a)--(f) show the vector fields of the lung deformations obtained from the registration results. The images are the slides on the x-axis with slide numbers $50,55,60,65,70$ and $75$ respectively. The vector fields located inside the lung are projected to the YZ planes and are visualized as green arrows in the figures. The vector fields are smooth, showing that our proposed algorithm can produce smooth landmark-based registration result.

{\color{red}

\begin{table}
\scriptsize
\begin{center}
\begin{tabular}{| c || c | c c c | c c c |}
\hline
& & \multicolumn{3}{c|}{Proposed} 		& \multicolumn{3}{c|}{Thin Plate Spline}	\\
\cline{2-8}
\multirow{2}{*}{}
& $\text{LM}_{\text{max}}$	& max K 	& min Det 	& $\text{e}_{\text{max}} / \text{e}_{\text{mean}}$ 	& max K 	& min Det	& $\text{e}_{\text{max}} / \text{e}_{\text{mean}}$ 	\\
&  $\text{LM}_{\text{mean}}$  	& \#LM  	& Time (s) 	& \#Fold 								& \#LM 	& Time (s) 	& \#Fold  							\\
\hline 
\multirow{2}{*}{One-point}
& 0.5196		& \cellcolor{Gray} 8.9756 	& \cellcolor{Gray} 0.2519		 	& \cellcolor{Gray}  0 / 0 	& \cellcolor{Gray} $\infty $ 	& \cellcolor{Gray} -0.0033 	& \cellcolor{Gray} 0.0017 / 0.0017 	\\
& 0.5196 		& 1 					& 11.3644 s				& 0				& 1 					& 0.08727 s  			& 2 						\\
\hline
\multirow{2}{*}{Two-point}
& 0.6164		& \cellcolor{Gray} 2.9295	 	& \cellcolor{Gray} 0.0989	 	& \cellcolor{Gray}  0 / 0 	& \cellcolor{Gray} $\infty$ 	& \cellcolor{Gray} -0.0722 	& \cellcolor{Gray} 0.0383 / 0.0352 	\\
& 0.4953 		& 2 					& 11.7554 s 				& 0				& 2 					& 0.0134 s 				& 213 						\\
\hline
\multirow{2}{*}{Twist}
& 0.5194		& \cellcolor{Gray} 4.1488	& \cellcolor{Gray}  0.1753 		& \cellcolor{Gray}  0 / 0	& \cellcolor{Gray} $\infty$ 	& \cellcolor{Gray} -0.0848		& \cellcolor{Gray} 0.0007 / 0.0004 	\\
& 0.3383  		& 50 					& 18.2556 s 				& 0				& 50 					& 0.0810 s 				& 131		 				\\
\hline
\multirow{2}{*}{Rotate}
& 0.5097		& \cellcolor{Gray} 3.1939	 	& \cellcolor{Gray} 0.4006 		& \cellcolor{Gray}  0 / 0 	& \cellcolor{Gray} $\infty$ 	& \cellcolor{Gray} -0.0755 	& \cellcolor{Gray} 0.0008 / 0.0004 	\\
& 0.3071 		& 3743 				& 28.7540 s 	 			&  0				& 3743 				& 61.0223 s 				& 35337 					\\
\hline
\multirow{2}{*}{Wave-shape}
& 0.2000		& \cellcolor{Gray} 3.3007	 	& \cellcolor{Gray} 0.3642	 	& \cellcolor{Gray}  0 / 0 	& \cellcolor{Gray} $\infty$ 	& \cellcolor{Gray} -0.0682 	& \cellcolor{Gray} 0.0007 / 0.0005 	\\
& 0.1256 		& 1089 				& 30.0948 s 				&  0				& 1089 				& 2.7450 s 				& 4603 					\\
\hline
\multirow{2}{*}{CT1}
& 0.0631		& \cellcolor{Gray} 2.2204	 	& \cellcolor{Gray}  0.1606		& \cellcolor{Gray}  0 / 0 	& \cellcolor{Gray} 1.0640 		& \cellcolor{Gray} 0.0085  	& \cellcolor{Gray} 0.0121/0.0074 	\\
& 0.0153 		& 300 					& 112.3901 s 			&  0				& 300 					& 1.3031 s  				& 0						\\
\hline
\multirow{2}{*}{CT2}
& 0.0624		& \cellcolor{Gray} 2.3331	 	& \cellcolor{Gray} 0.1406 		& \cellcolor{Gray}  0 / 0 	& \cellcolor{Gray} 1.0740 		& \cellcolor{Gray} 0.0843  	& \cellcolor{Gray} 0.0136/0.0077 	\\
& 0.0263 		& 300 					& 103.3852 s 			&  0				& 300 					& 1.1630 s 				& 0	 					\\
\hline
\multirow{2}{*}{CT3}
& 0.0891		& \cellcolor{Gray}1.7137	 	& \cellcolor{Gray} 0.3973		& \cellcolor{Gray}  0 / 0 	& \cellcolor{Gray} 1.1763 		& \cellcolor{Gray} 0.0593	 	& \cellcolor{Gray} 0.0123 / 0.0077 	\\
& 0.0314 		& 300 					& 90.1452 s 				& 0				& 300 					& 1.1706 s 				& 0						\\
\hline
\multirow{2}{*}{CT4} 
& 0.0816		& \cellcolor{Gray} 6.1340	 & \cellcolor{Gray} 0.0312			& \cellcolor{Gray}  0 / 0 	& \cellcolor{Gray} 1.3528		& \cellcolor{Gray} 0.0639	 	& \cellcolor{Gray} 0.0138 / 0.0079 	\\
& 0.0393 		& 300 					 &  81.7555 s 			&  0		 		& 300 					& 1.2033 s 				& 0 	 					\\
\hline
\multirow{2}{*}{CT5}
& 0.0920		& \cellcolor{Gray} 6.5297	& \cellcolor{Gray} 0.0229			& \cellcolor{Gray}  0 / 0 	& \cellcolor{Gray} 1.3226 		& \cellcolor{Gray} 0.0384		& \cellcolor{Gray} 0.0159 / 0.0073 	\\
& 0.0232 		& 300 					&  228.0194 s  			& 0				& 300 					& 1.5251 s 				& 0	 					\\
\hline
\end{tabular}
\par\end{center}
\caption{Quantitative measures of the registration experiment.\label{tab:Registration_result}}
\end{table}

}

\subsection*{Quantitative measurements}

Table \ref{tab:Registration_result} lists the quantitative measurements of the landmark-matching transformation obtained from the proposed algorithm and the TPS method. For a fair comparison, we first normalize the domain $\Omega$ in each example to be the unit cube. The maximum and minimum displacement of the prescribed landmark correspondences are denoted as $\text{LM}_{\max}$ and $\text{LM}_{\min}$ respectively. The quantities $e_{\max}$ and $e_{\min}$ shows the maximum and minimum landmark mismatching error in the L-2 sense. 

The maximum of the resulting conformality distortion is denoted by max K. Note that max K obtained from our proposed algorithm are all finite. This implies that the computed transformations in all examples are orientation-preserving. However, results generated by the TPS in the five synthetic examples have infinite value of max K, which indicates folding occurs in the mapping obtained from TPS. For lung registration examples (CT1 - CT5), we observe that the max K of TPS is relatively smaller than that of the proposed algorithm. This is mainly due to the inexact alignment of the landmark points by TPS which provides more freedom for the optimization of the transformation. $\min \text{Det}$, which is the minimum of the Jacobian, is another indicator showing the diffeomorphic property of the mapping \cite{christensen2001consistent,christensen1996deformable}. $\#\text{Fold}$ counts the number of tetrahedra in which the obtained transformation has negative Jacobian. We observe that foldings occur in TPS method when the landmark displacement is large. For our proposed algorithm, no foldings are observed for both synthetic and the lung registration examples. This shows the capability of the generalized conformality distortion $K(f)$ in enforcing the bijectivity of the transformation. 


The computation time for both algorithms is also reported in the table. With the proposed numerical method applied in the algorithm, the time required for large deformation is quite reasonable (less than 30 seconds for sparse grids and less than four minutes for dense grid size).

\section{Conclusion}
\label{sec:7}
This paper present a new method to obtain folding-free landmark-matching transformationn between general $n$-dimensional Euclidean spaces with large deformations. The basic idea is to extend the 2-dimensional quasi-conformal theories to general $n$-dimensional spaces. Given a set of landmark constraints, our goal is to look for an optimal transformation that matches landmarks. In this paper, we introduce a notion of conformality distortion of a diffeomorphism of the $n$-dimensional Euclidean space. The conformality distortion measures the distortion of an infinitesimal ball to an infinitesimal ellipsoid under the diffeomorphism. Our problem can then be modelled as a minimization problem of an energy functional involving the conformality term and a smoothness term. The conformality term allows the algorithm to produce folding-free transformation with minimized local geometric distortions, even with very large deformations. Alternating direction method of multipliers (ADMM) is applied in this paper to solve the optimization problem. The algorithm only involves solving an elliptic problem and a tetrahedron-wise minimization problem. Preconditioned conjugate gradient method with multi-grid V-cycle preconditioner is applied to one of the subproblem, while a fixed-point iteration is used for another subproblem. The time complexity and robustness of the algorithm is independent of the number of landmark constraints. Experimental results show that our proposed algorithm is effective for computing folding-free landmark-matching transformation, even with large number of landmarks or large deformations. In the future, we will test the algorithm on other real medical data, such as 3D MRI scan with DTI fibre tracks as the interior landmark constraints.

\begin{acknowledgements}
The authors acknowledge the freely available lung CT data from the Deformable Image Registration Laboratory (\url{www.dir-lab.com}).
\end{acknowledgements}

\bibliography{reference}
\bibliographystyle{plain}

\end{document}